# Theoretical Foundations of Community Rating by a Private Monopolist Insurer: Framework, Regulation, and Numerical Analysis

Yann Braouezec[a,*], John Cagnol[b,c,*]

[a] *IESEG School of Management, Univ. Lille, CNRS, UMR 9221 - LEM - Lille Economie Management, F-59000 Lille, France, y.braouezec@ieseg.fr.*

[b] *Mathematics and Computer Science Laboratory for Complexity and Systems, CentraleSupélec University Paris-Saclay, 3 rue Joliot Curie, 91190 Gif-sur-Yvette, France*

[c] *Federation of Mathematics of CentraleSupélec, CNRS FR3487*

**Abstract**

Community rating is a policy that mandates uniform premium regardless of the risk factors. In this paper, our focus narrows to the single contract interpretation wherein we establish a theoretical framework for community rating using Stiglitz's (1977) monopoly model in which there is a continuum of agents. We exhibit profitability conditions and show that, under mild regularity conditions, the optimal premium is unique and satisfies the inverse elasticity rule. Our numerical analysis, using realistic parameter values, reveals that under regulation, a 10% increase in indemnity is possible with minimal impact on other variables.

## 1. Introduction

Community rating is a regulatory principle that maintains uniformity in insurance premiums across a given geographic area, thereby prohibiting insurers from adjusting premiums based on factors such as age, gender, health status, or other relevant characteristics[1]. In essence, community rating enforces an equal premium for all agents (policyholders) regardless of their individual risk profiles. The United States' Affordable Care Act (ACA), commonly referred to as Obamacare, introduced a modified form of community rating. This variant permits health insurers offering ACA-compliant plans to charge senior citizens premiums up to three times higher than those paid by younger enrollees. While often associated with health insurance [15, 41], this concept has attracted interest across a spectrum of insurance domains, including automobile, property, and casualty coverage.

Whether in its pure or modified form, community rating doesn't prohibit the creation of multiple insurance contracts featuring varying levels of coverage. In [29] and [31], the authors investigate a framework where an insurer offers two distinct contracts while adhering to community rating constraints. Each contract entails a uniform community-rated price, meaning that this price remains unaffected by risk characteristics like age, gender, or health status. However, within a monopoly context, where the insurer has leeway in determining the number of contracts, even under circumstances of adverse selection, there exists the possibility of constructing an appropriate menu of contracts so that each policyholder pays a premium

---

[*] Equally contributing authors

*Email addresses:* `y.braouezec@ieseg.fr` (Yann Braouezec), `john.cagnol@centralesupelec.fr` (John Cagnol)

[1] For a comprehensive definition, refer to https://www.healthcare.gov/glossary/community-rating/

related to their individual risk attributes, as explored in works such as [46] and [18]. This is in sharp contradiction with the spirit of community rating, even in its modified form. Essentially, within the context of a monopoly, the implementation of community rating, particularly in its pure form, implies the necessity of regulating the number of contracts offered. This translates to a constraint on the insurer, requiring them to provide only a single insurance contract to the entire community. Consequently, the core concept of pure community rating equates to the absence of any form of direct or indirect price discrimination. This premise resonates with discussions found in the literature on third-degree price discrimination in monopolies, which seeks to determine the conditions under which price discrimination contributes positively to social welfare. For more insights, see [53], and for a broader overview, refer to [54], see also [9] for model in which the number of market segments is endogenous.

The present paper aims to establish the theoretical foundations of a community rating insurance system, focusing specifically on the interpretation involving a single contract. Building upon the foundation laid in [46] (also explored in [18], [32]), we work with a *continuum* of agent types distributed according to a probability measure, which may or may not exhibit a continuous density. In this setting, insurance remains non-mandatory. As in [32] but as opposed to [46] or [18], we allow for a positive fraction of agents to possess an extreme probability of damage, thereby characterizing the set of types as a unit compact within the real line. The insurer faces an adverse selection problem wherein the type of a given agent remains unobservable, though the distribution of types (as represented by the probability measure) is known. Following the premise of [46], we assume that heterogeneity among agents solely stem from their types, with the monopolist insurer possessing comprehensive knowledge except for identifying the type of each agent, i.e., the probability of damage. Our deviation from [46] lies solely in the imposition of a constraint upon the profit-maximizing monopolist insurer, requiring them to furnish a single insurance contract. This framework, characterized by a *continuum* of agents, holds dual significance. Firstly, it captures the notion of a large number of agents, allowing the application of the law of large numbers [52]. Secondly, the segmentation of agent subsets (insured versus uninsured) emerges as an endogenous facet, subject to characterization under the condition of regularity in the underlying probability measure (continuously differentiable). To the best of our knowledge, such a single contract underpinning (pure) community rating has received minimal attention within the theoretical economics literature[2] and is the focus of the present study. From a pure theoretical point of view, our paper can be seen as a natural complement to [46] and [18] since we consider the same framework but under different regulatory regime. [46] and [18] analyze the fully unregulated case while study the regulated one, i.e., community rating.

Our analysis investigates the scenario of a private monopolist insurer mandated to provide a singular insurance contract. Within this framework, the single insurance contract can be characterized by a pair of positive parameters: the indemnity ($R$), denoting the contract's quality, and the premium ($P$), reflecting its price. This contract is presented to the overall set of agents (the community) on a "take it or leave it" basis. The premium is remitted by agents at the inception of the contract to secure policyholder status over a unit period. In return, the insurer disburses indemnities to policyholders who incur damages during the contract's defined unit period, typically a year.

From a regulatory standpoint, the endogenous nature of both quality and price naturally calls for their optimization to maximize social welfare, following the insights of [45] and [44]. Analogous to observations made in [26] and related works, the insurer finds themself in a state of natural monopoly, since the average cost is always above the marginal cost. As social welfare maximization mandates the adoption of marginal cost pricing for each indemnity level, the first-best approach entails an inherent deficit. Consequently, the pursuit of a second-best pricing rule is necessitated, as explored in [14], [35], and [43]. A plausible regulatory strategy could involve compelling the monopolist insurer to extend an insurance contract of utmost quality (akin to full coverage) while adhering to profitability constraints. The premium for such comprehensive coverage might prove considerable, potentially resulting in a less-than-satisfactory take-up rate. Alternatively, a regulatory avenue could simply stipulate a minimum contract quality threshold, say 60% of the loss, akin to the limited coverage examined in [31]. In our approach, we adopt the following regulatory approach for the scenario at hand: we assume that, given the indemnity $R$, the insurer designs the market segment (those who become policyholders) to maximize its profits. This optimal segmentation, contingent on $R$, then guides the regulator's decision on $R$, aiming to optimize social welfare[3]. Subsequently, this regulated scenario is compared to the non-regulated counterpart in terms of premiums, indemnities, and take-up rates.

---

[2]Contrastingly, the empirical landscape pertaining to community rating is considerably expansive, evident in works such as [15], [1], and [41], among others. See also [8].

To comprehensively analyze the regulated scenario, where the indemnity $R$ is predefined, a profitability assessment of the market must be performed. In the model established by [46], the insurer's information asymmetry only pertains to the agent's type, enabling the computation of an agent's willingness-to-pay, denoted as $\overline{P}(\theta, R)$, when the agent possesses a specific type $\theta$. Here, when the premium $P$ chosen by the insurer matches $\overline{P}(\theta, R)$ for a given $\theta$, only agents with greater willingness to pay opt for the contract. By choosing $\theta$, the insurer effectively delineates the market segment $[\theta, 1]$ to encompass policyholders. Within the regulated framework, where $R$ is predetermined, the insurer's problem of profit maximization simplifies into a one-dimensional challenge similar to the classical monopolist problem, although more complex. We present a sufficient condition for the existence of a profitable contract, and it turns out that this condition involves the derivative of the *average probability of damage* within the group $[\theta, 1]$, evaluated at the highest type, i.e. 1, denoted as $A'(1)$. While the average probability of damage remains undefined at 1, we unveil conditions that render this derivative well-defined in 1. Assuming the fulfillment of this profitability criterion, we demonstrate that the optimal premium satisfies the Lerner index principle. In other words, it aligns with the conventional inverse elasticity rule widely discussed in the microeconomics literature (e.g., [55]). Compared to the classical monopoly theory, the elasticity incorporates an additional derivative component due to the endogenous nature of both premium and quantity variables, i.e., only the segmentation is exogenous. Interestingly, this elasticity can be split into the product of two terms: one linked to the "statistical aspect" of the market (the hazard rate) and another associated with the "preference aspect" (the logarithmic derivative of the premium). At this stage, uniqueness of the optimal premium can not be guaranteed. However, we establish that in cases where the average probability of damage exhibits linearity or convexity with respect to type, the optimal premium indeed becomes unique. This outcome stands as a pivotal result within this paper.

Within the existing literature, our framework featuring a single contract and a *continuum* of agents has exclusively been examined in instances where the underlying probability measure takes the form of a Dirac mass. In such cases, it is demonstrated that the profit-maximizing contract translates to full coverage, as shown in works such as [46], with further insights available in [23]. However, this Dirac mass framework lacks interest, as it assumes agents to be entirely homogeneous in terms of risk. Essentially, it mimics the scenario of a solitary agent possessing a singular probability of damage. Given the known probability measure, the insurer can deduce this singular agent's type, thus defeating adverse selection. With only one type in play, the insurer can capture all the policyholders' surplus by offering a contract where the premium equals the unique agent's willingness to pay. Consequently, the distinction between our two regulatory scenarios (regulation of indemnity or its absence) becomes inconsequential, as social welfare inevitably equates to total profit. However, when the underlying probability measure admits a continuous density, the insurer loses the ability to wholly extract the policyholders' surplus. Consequently, uncertainty prevails over whether the two regulatory scenarios remain identical or diverge. We delve into both cases: the unregulated instance where the monopolist insurer wields the freedom to determine segmentation and indemnity, and the regulated scenario where the regulator dictates the indemnity. No theoretical result however can be offered regarding the solution of the social welfare maximization. To assess the potential value of indemnity regulation, we present a numerical simulation where we consider a Constant Absolute Risk Aversion (CARA) utility function, and the type distribution adheres to a Beta density. Following the methodology outlined in a companion paper, [12], we perform a change of variables to confine the parameters of the Beta distribution within a bounded subset of $\mathbb{R}_+^2$, precisely, its unit square. This transformation facilitates a comprehensive analysis. From an economic perspective, we opt for parameter values that reflect typical values found in the literature, drawing upon works such as [4]. Through this approach, we explore the efficacy of regulation. Our numerical findings show that, particularly when relative risk aversion is higher than 5, indemnity regulation yields substantial benefits. In specific terms, policyholders are empowered to avail themselves of significantly enhanced indemnities exceeding 11% while experiencing a marginal increase in premiums of less than 2%. The code employed for this study is accessible in [11].

---

[3]This regulatory mechanism assumes that the regulator possesses the same information and capabilities as the insurer.

The structure of this paper is outlined as follows. The first part of the paper (comprising Sections 3, 4, and 6) explores the pure community approach within the framework of the [46] model. Here, the insurer is constrained to offer a single insurance contract (defined by indemnity and premium), and we thoroughly examine both the regulated and unregulated cases. The second part of the paper (Section 7) offers a comprehensive numerical analysis. Most of the proofs are relegated in the appendix.

## 2. Brief Literature Review

In this section, we provide a concise overview of theoretical papers within a monopoly context that bear direct relevance to our present study. Notably, we avoid delving into the extensive empirical literature concerning community rating, nor do we delve into the theoretical literature centered around insurers competing for instance à la Bertrand, as in [27] and [31]. See also [56] for a recent paper devoted to insurance regulation in a competitive framework. To enhance clarity, we follow the presentation approach outlined in [7]. Our discussion initially explores the scenario where the monopolist insurer operates without regulatory constraints, followed by the examination of situations where some form of regulation is applied. Furthermore, we point readers to [3] for a brief survey on log-concavity's application in monopoly theory, albeit with no direct connection to insurance.

*Monopoly Solution without Regulation.* In this context, the monopolist insurer possesses the freedom to design a menu of contracts, albeit limited by its informational constraints, i.e., it is limited by its knowledge of the agents' risk attributes. In an idealized setting of complete information, where an agent's characteristics are entirely known, the optimal contract menu simplifies: each agent receives a comprehensive coverage insurance contract with a premium equivalent to their willingness-to-pay. This approach allows the insurer to capture the entirety of the surplus (refer to [10] for a complete information model and its justification). In scenarios involving adverse selection, wherein the insurer lacks perfect knowledge about each agent's risk attributes, a contract menu is devised to extract as much as possible of the insured surplus. In his seminal paper, [46] (see [18], [32], [42], [48] for closely related models) offers a model along this idea; the insurer faces a *continuum* of agents (i.e., of types) and can freely choose the menu of contracts. He shows that this optimal menu, depending upon conditions, may entail pooling (a situation in which two agents with different risk characteristics buy the same contract) or complete separation (called complete sorting in [18]), a situation in which contracts are perfectly customized. The framework re-evaluated by [18] presents a rigorous mathematical analysis and introduces additional insights. For instance, they demonstrate that complete sorting doesn't necessarily imply every agent becomes insured (when agents are sufficiently risk-adverse, a fraction may remain uninsured). A striking feature of [46] and [18] is that the unique cost considered is the cost of claims. In particular, there are no administrative costs, that is, no cost of claim processing and no issuance cost ([22]). Recently, a number of authors have introduced these administrative costs in their analysis, e.g., [13], [22], [19]. As noted in [22], these administrative costs obviously raise the cost functions and when they are too high, the market is not anymore profitable, i.e., there are no gain to trade. In [19] (see Proposition 1), they provide a set of conditions under which there are no gain to trade in the case of "smooth cost". In [13], they offer an interesting but fairly complex health insurance model with administrative cost in the popular Rothschild-Stiglitz two types framework. While there are no clear theoretical results, the valuable aspect of their model is that it is calibrated on data.

*Monopoly Regulation.* Monopolist insurers can be subject to diverse regulatory mechanisms. Beyond solvency ratios requiring risk-based capital ratios to exceed a regulatory threshold, pricing of insurance contracts (namely the rate or premium) is a crucial regulatory focus. From a regulatory standpoint, as highlighted in [40] (refer to Chapter 8 on insurance regulation), these rates must avoid both excessive charges and unfair discrimination, while maintaining the solvency of insurance groups. Several jurisdictions prohibit the use of specific risk characteristics like age or gender in auto insurance contracts, or genetic testing in life insurance (see [24], and [17] for recent reviews). As discussed in the introduction, pure community rating mandates uniform rates for a given plan across all individuals or small groups, regardless of age, sex, or other risk attributes [15]. Community rating has found substantial application in health insurance (e.g., [15], [41], [8], [20]), and it extends to automobile insurance (e.g., [51]).

## 3. A General Model of Community Rating

To analyze community rating, we shall use the framework established by [46], which involves a *continuum* of agents distributed according to a probability measure. As opposed to [46], where the insurer operates without regulation, we will assume a regulated scenario and will impose the insurer offers a single contract, akin to the notion of pure community rating. However, we will not confine the model to a specific domain such as health insurance, casualty property insurance, and the like, as the framework remains universally applicable across diverse insurance sectors.

*3.1. Assumptions and Discussion*

Following [46], we consider the simplest framework to work with when there are infinitely many agents. Each agent is differentiated with respect to the type $\theta \in [0,1]$, which is simply the probability of damage of a given agent. By abuse of notation, we shall call a given agent by her type, that is, agent $\theta$ is the agent with a type equal to $\theta$. Let $[0,1]$ be the set of types (or agents) distributed according to a probability measure $\mu$. At this stage, $\mu$ is a general probability measure and needs not admit a density but we will eventually consider three levels of hypotheses described in forthcoming Definition 4.

Risk-averse agents are endowed with the same initial wealth $W_0 > 0$. In case of a damage, the loss is equal to the constant $L$, where $L \in (0, W_0]$. Both $W_0$ and $L$ are dimensional variables with the dimension of a currency unit. Their non-dimensional counterparts will be introduced in Section 7. Introducing stochastic loss $L$ is straightforward. Conditional upon a given damage, if $L$ is identically and independently distributed, under the conventional assumption of insurer risk-neutrality, the replacement of $L$ with its expectation $\mathbb{E}(L)$ suffices. Since risk-averse agents are differentiated solely by their type, they share a uniform Bernoulli utility function denoted as $W \mapsto U(W)$, where $W \geq 0$ is the wealth. This utility function is assumed to be twice continuously differentiable. As usual in Economics, we also assume that $U$ is strictly increasing and concave since agents are risk-averse. Some stronger results will also be established when $U''$ is increasing as specified later, although this is not a requirement for most of the results. As in [46], the insurer knows the utility function $U$, the initial wealth $W_0$ and the loss $L$ in case of damage. Only the type $\theta$ is unknown to the insurer, that is, the insurer faces an adverse selection problem. As opposed to [46] and subsequent literature (e.g., [18], [32]), we make the assumption that under pure community rating, the monopolist insurer is constrained to offer a *single* insurance contract $C$, that is, the insurer cannot offer more than one contract.

Consider the final wealth of agent $\theta \in [0,1]$, which is the random variable equal to $W_0 - L$ with probability $\theta$ and to $W_0$ with probability $1 - \theta$. Since $[0,1]$ is a non-empty interval of $\mathbb{R}$, it has the cardinality of the *continuum*, which means that we work with a *continuum* of independent Bernoulli random variables, a framework similar to [34] (see also [52]). In contrast to various insurance models featuring a *continuum* of agents ([18], [42], [46], [48]) where the space of types is typically confined within $(0,1)$, we adopt a more general approach and do not impose a bound on types away from one. Thus, the types considered span the space $[0,1]$, as in [32].

Each risk-averse agent is assumed to assess the alternatives based on the expected utility of their final wealth. Consider first the case without an insurance company. As said earlier, the final wealth of the agent $\theta$ is the random variable $X_\theta$ so that her expected utility of the final wealth denoted $V(\theta, 0)$ is equal to

$$V(\theta, 0) := \theta U(W_0 - L) + (1 - \theta) U(W_0) \tag{1}$$

and is a decreasing function of $\theta$.

Consider now the case with an insurance company that offers a single non-mandatory contract. A contract is defined by a premium $P \in [0, L]$ and an indemnity $R \in [0, L]$ to be paid in case of damage, both variables are dimensional variables with the dimension of a currency unit. In line with the legal requirement in several states of USA (over-insurance is prohibited in 15 states) and countries such as France, we shall assume that the indemnity cannot exceed the loss, that is $R \leq L$. As usual in the literature, when $R = L$, such a contract is called a *full coverage contract* while when $R < L$, it is called a *partial coverage contract*. We shall note this contract $C = (P, R) \in [0, L] \times [0, L]$.

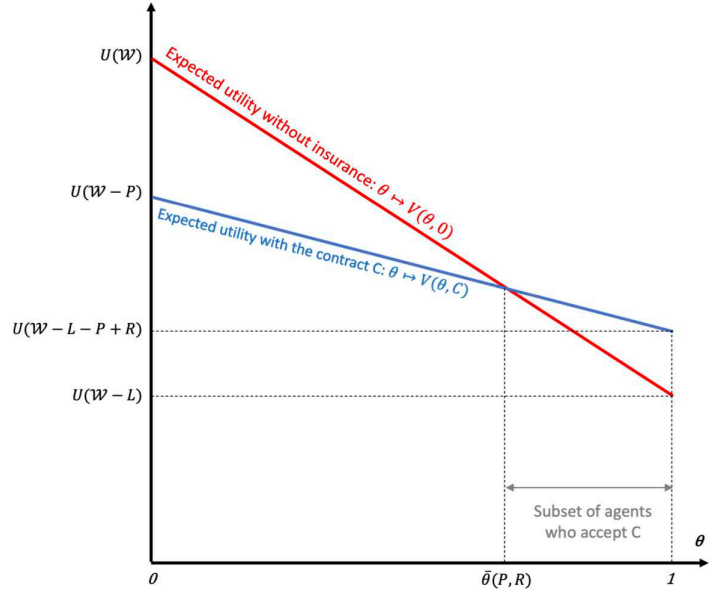

Figure 1: Participation is satisfied when $C$ is admissible

When the contract $C$ is purchased by agent $\theta$, her final wealth is a random variable that takes two values, $W_0 + R - L - P$ with probability $\theta$ and $W_0 - P$ with the complementary probability $1 - \theta$. The expected utility of the final wealth of agent $\theta$ denoted $V(\theta, C)$ thus is equal to

$$V(\theta, C) := \theta U(W_0 + R - L - P) + (1 - \theta)U(W_0 - P) \qquad (2)$$

and note that $R - L$ can be interpreted as a deductible which is beared by the agent. The absence of insurance is represented by the contract $C = (0, 0)$, conventionally denoted as 0. This explains the introduction of the notation $V(\theta, 0)$ in equation (1), which signifies the expected utility of the final wealth without insurance.

In the context of complete information, wherein the type of each agent is observable, the insurer can offer the contract to a selected group of agents. This method entails a *direct* segmentation, where the decision to insure an agent rests with the insurer (see, e.g. [10] for a model incorporating complete information within a *continuum* of agents). However, this dynamic takes a different course in our situation where the type is unobservable. Here, the insurance company designs a contract $C$, yet remains powerless to prevent any agent from acquiring it. The segmentation of the set of agents is regarded as *indirect*, since the decision to become a policyholder lies within the discretion of each individual agent [36]. Consider a situation where the insurance contract $C$ offered. Given that insurance remains non-compulsory, an agent will opt to purchase the contract $C$ if her expected utility, associated with the insurance contract, surpasses or equals her expected utility without insurance, that is, if

$$V(\theta, C) \geq V(\theta, 0) \qquad (3)$$

As a usual tie-breaking assumption, when an agent $\theta$ is indifferent between the two situations (i.e., to become insured or not), she chooses to be insured.

**Fact 1.** *If a given agent $\theta_0 \in (0, 1)$ purchases the contract $C$, then, all agents such that $\theta \in [\theta_0, 1]$ also purchase the contract $C$.*

The proof is straightforward, stemming from the simple observation that $\theta \mapsto V(\theta, C) - V(\theta, 0)$ is a monotonically increasing function. Within our model, we shall say that a contract $C$ fulfills the participation constraint if there exists $\theta \in [0, 1)$ such that $V(\theta, C) \geq V(\theta, 0)$. In this case, we say the contract contract is deemed *admissible*. The set of all admissible contracts is denoted as $\mathcal{C}_\mathcal{A}$ and is defined as follows.

$$\mathcal{C}_\mathcal{A} := \{C \in [0, L] \times [0, L], \; \exists \theta \in [0, 1) \mid V(\theta, C) \geq V(\theta, 0)\} \tag{4}$$

As long as $C \in \mathcal{C}_\mathcal{A}$, there exists an agent $\theta < 1$ who purchases this contract $C$ and the participation constraint is fulfilled. From Fact 1, the set of agents who purchase the contract (the set of policy holders) is a non-empty market segment of the form $[\theta, 1]$. The following lemma, proved in the appendix, establishes a simple characterization of the set of admissible contracts $\mathcal{C}_\mathcal{A}$.

**Lemma 1.** *The set $\mathcal{C}_\mathcal{A}$ of admissible contracts is characterized by*

$$C = (P, R) \in \mathcal{C}_\mathcal{A} \iff P < R$$

Note that we require $\theta$ to belongs to $[0, 1)$ and not to $[0, 1]$ in (4). The latter is always satisfied (since agent $\theta = 1$ always purchases the contract) and is insufficient to derive Lemma 1.

*3.2. Critical Thresholds*

In our model, the insurer chooses three variables: $\theta$, $P$, and $R$. We note $C = (0, 0)$ corresponds to a cost-free insurance contract with zero indemnity. Using the notation introduced before, the expected utility without insurance is denoted as $V(\theta, 0)$. Let us define

$$\begin{aligned}\mathcal{G}(\theta, P, R) &:= V(\theta, P, R) - V(\theta, 0) \\ &= \theta U(W_0 - L - P + R) + (1 - \theta)U(W_0 - P) - \theta U(W_0 - L) - (1 - \theta)U(W_0)\end{aligned} \tag{5}$$

By construction, $\mathcal{G}(\theta, P, R) \geq 0$ (respectively negative) means that agent $\theta$ accepts (respectively rejects) the contract. Because $U$ is twice continuously differentiable, so is the function $\mathcal{G}$. From the Implicit Function Theorem, there exist *critical threshold functions* of class $\mathcal{C}^2$

$$\bar{\theta} : (0, L) \times (0, L) \to (0, 1) \quad \text{such that} \quad \forall (P, R) \in (0, L) \times (0, L), \; \mathcal{G}(\bar{\theta}(P, R), P, R) = 0 \tag{6a}$$
$$\bar{P} : (0, 1) \times (0, L) \to (0, L) \quad \text{such that} \quad \forall (\theta, R) \in (0, 1) \times (0, L), \; \mathcal{G}(\theta, \bar{P}(\theta, R), R) = 0 \tag{6b}$$
$$\bar{R} : (0, 1) \times (0, L) \to (0, L) \quad \text{such that} \quad \forall (\theta, P) \in (0, 1) \times (0, L), \; \mathcal{G}(\theta, P, \bar{R}(\theta, P)) = 0 \tag{6c}$$

This phenomenon is encapsulated succinctly in this fact:

**Fact 2.** *The choice of two decision-variables determines the third one.*

Consider first the natural situation in which the insurer both chooses $P$ and $R$. Then, $\theta$ is implied by this choice and is given by equation (6a). Solving $\mathcal{G}(\theta, P, R) = 0$, the implied function (6a) can indeed be given explicitly:

$$\bar{\theta}(P, R) = \frac{U(W_0) - U(W_0 - P)}{[U(W_0) - U(W_0 - P)] + [U(W_0 - L + R - P) - U(W_0 - L)]} < 1 \tag{7}$$

When the insurer designs an admissible contract $C = (P, R)$, the agents with a type $\theta$ greater or equal to $\bar{\theta}(P, R)$ will opt to purchase the contract, which implies that the group of policyholders is the interval $[\bar{\theta}(P, R), 1]$, as illustrated in Figure 1. It is thus natural to call the function $\bar{\theta}(P, R)$ the *critical agent* or *critical threshold* for $C$. It is a non-dimensional quantity within the range of $[0, 1]$. The critical agent obviously depends upon $P$ and $R$ but also upon her risk-aversion through the utility function as shown in (7).

Alternatively, the insurer could chose the indemnity $R$ and $\theta$ in such a way that the premium becomes the implied function, which is the maximum price the agent $\theta$ is willing to pay for the insurance contract with indemnity $R$. Using the classical terminology, $\bar{P}(\theta, R)$ is simply the willingness-to-pay (WTP) of agent $\theta$. Given that $\bar{P}(\theta, R)$ increases with $\theta$, when the premium equals $\bar{P}(\theta, R)$, any agent $\theta'$ where $\theta' > \theta$ will purchase the contract. In instances where the insurer choses $R > 0$ and $\theta$, the resulting contract becomes $C = (\bar{P}(\theta, R), R)$, with the set of policyholders is $G = [\theta, 1]$.

**Definition 1.** *Let $\theta \in (0,1)$ and $R \in [0, L]$. The function $\bar{P}(\theta, R)$ is a critical premium called the reservation price or the maximum willingness-to-pay. It lies in $[0, L]$ and its dimension is in a currency unit.*

Following this, the selection of $\theta \in (0,1)$ and $R \in [0, L]$ produces a segmentation of the agent set $[0, 1]$ into two distinct categories: the policyholders ($G = [\theta, 1]$), who chose to purchase the insurance, and those who remain uninsured ($\bar{G} = [0, \theta)$). Given that the choice of $\theta$ results in such a partition of agents, distinguishing between those with coverage and those without, it naturally assumes the role of a *market segmentation variable*.

*3.3. Willingness-to-Pay for the Insurance Contract, Profit, and Social Welfare*

For a given choice of $\theta$ and $R$, the contract offered is $C = (\bar{P}(\theta, R), R)$, and the group of policyholders is the market segment $G = [\theta, 1]$. The following proposition, proved in the appendix, outlines several properties of the willingness-to-pay $\bar{P}(\theta, R)$.

**Proposition 1.** *Let $R \in [0, L]$ and $\theta \in (0, 1)$.*

(i) *We have $\bar{P}(0, R) = 0$ and $\bar{P}(1, R) = R$.*

(ii) *The functions $\theta \mapsto \bar{P}(\theta, R)$ and $R \mapsto \bar{P}(\theta, R)$ are increasing.*

(iii) *If $U''$ is increasing or $R = L$, then $\theta \mapsto \bar{P}(\theta, R)$ is concave and $\bar{P}(\theta, R) \geq \theta R$.*

In the case of a full coverage contract ($R = L$), the maximum willingness-to-pay can be explicitly given by $\bar{P}(\theta, L) = W_0 - U^{-1}(V(\theta, 0))$. This comes from the fact that each policyholder's final wealth is invariably $W_0 - \bar{P}(\theta, L)$, and her utility is $U(W_0 - \bar{P}(\theta, L))$. This situation was explored in [10], revealing that the maximum willingness-to-pay is an increasing and concave function of $\theta$. As a result, the aforementioned proposition extends this finding to the case where $R \leq L$. Notably, it should be noted that the function $R \mapsto \bar{P}(\theta, R)$ is generally increasing but not concave.

In the case of a CARA utility function $U$ given by $U(W) = 1 - e^{-\lambda W}$, where $W \geq 0$ represents the wealth and $\lambda > 0$ denotes the Arrow-Pratt risk aversion parameter, the requirement for $U''$ to be increasing is automatically satisfied. The willingness-to-pay denoted $\bar{P}_\lambda(\theta, R)$ can be computed explicitly by solving the equation (6b).

$$\bar{P}_\lambda(\theta, R) = \frac{1}{\lambda} \times \ln\left(\frac{\theta e^{\lambda L} + (1-\theta)}{\theta e^{\lambda(L-R)} + (1-\theta)}\right) \tag{8}$$

**Fact 3.** *For the CARA utility function, the willingness-to-pay $\bar{P}_\lambda(\theta, R)$ is concave in $\theta$ and $\bar{P}_\lambda(\theta, R)$ concave[4] in $R$.*

For a given $R \in [0, L]$ and $\theta \in (0, 1)$, let $C = (\bar{P}(\theta, R), R)$ be the contract offered to the overall set of agents. By construction, only the group of agents $[\theta, 1]$ purchase the contract. Consider now an agent with type $x \in [\theta, 1]$. The (expected) profit of the insurer derived from the purchase of the contract $C$ by agent $x$ is given by the premium minus the (expected) cost of claims:

$$\bar{P}(\theta, R) - xR$$

---

[4]It should be pointed out that the concavity in relation to $R$ doesn't hold in general.

By integration over the entire market segment $[\theta, 1]$, the total profit of the insurer, denominated in a given currency, is equal to

$$\Pi(\theta, R) = \int_\theta^1 (\bar{P}(\theta, R) - xR)\, d\mu(x) = \underbrace{\bar{P}(\theta, R)\mu([\theta, 1])}_{\text{Total Revenue}} - \underbrace{R \int_\theta^1 x\, d\mu(x)}_{\text{Total Cost}} \qquad (9)$$

**Remark 1.** *In studies such as [22] and [19], the consideration of the cost of claim processing, denoted as $c_1 > 0$ in [22], is also explored. When a policyholder with a given type $x$ experiences a loss, the insurer faces an additional cost related to the processing of the claim for this agent $x$. This cost is incurred only in cases of actual damage. Consequently, when the insurer offers a contract $C = (P, R)$, the expected cost for the insurer associated with agent $x$ becomes $x(R + c_1)$ instead of $xR$, reflecting the inclusion of the processing cost. As a result, the total profit of the insurer can be represented as*

$$\bar{P}(\theta, R)\mu([\theta, 1]) - (R + c_1) \int_\theta^1 x\, d\mu(x)$$

*In our model, the incorporation of the cost of claim processing (as in [22] and [19]) or the inclusion of a fixed cost (as in [10]) would have minimal impact on the overarching findings of this paper. For simplicity, we will assume without loss of generality that $c_1 = 0$. In the same vein, we make the implicit assumption that the total revenue from the collect of the premiums is invested in a risk-free asset whose interest rate $i$ is normalized to zero.*

The surplus of policy holders thus is equal to

$$\text{SP}(\theta, R) := \int_\theta^1 (\bar{P}(x, R) - \bar{P}(\theta, R))\, d\mu(x) \qquad (10)$$

and the social welfare is given by

$$\text{SW}(\theta, R) := \Pi(\theta, R) + \text{SP}(\theta, R) = \int_\theta^1 (\bar{P}(x, R) - Rx) d\mu(x) \qquad (11)$$

The social welfare is independent of the premium charged $\bar{P}(\theta, R)$ but is explicitly dependent on the selection of the market segmentation parameter $\theta$. In alignment with the terminology of [5], we shall adopt the assumption that the willingness to pay serves as a sufficient statistic for consumer welfare. For a more comprehensive discussion, readers can refer to their Subsection 2.2.

*3.4. Community Rating, Optimal Contract, and the Natural Monopoly Problem*

Within our community rating framework, characterized by a single contract, the decision variables can be chosen to be the market segmentation $\theta$ and the indemnity $R$. This gives direct access to the contract quality (i.e. the indemnity) and the percentage of the population which is covered ($\mu([\theta, 1])$), often called the take-up rate. The market could operate without regulatory intervention. In this case, the insurer will to maximize the profit $\Pi(\theta, R)$ and identify the optimal values $\theta^*$ and $R^*$. These optimized parameters subsequently establish the premium $P^* = \bar{P}(\theta^*, R^*)$.

On the other hand, the market can also be subject to regulation. Regulatory bodies may establish a "minimum contract quality" criterion, compelling insurers to provide contracts where $R$ surpasses a predetermined threshold. For example, a regulator might stipulate that $R$ must be at least 80% of $L$. Alternatively, regulators could enforce a requirement that the insurer covers a specific fraction of the agent population, such as 50%. In this scenario, the regulator would set $\theta$ to be no greater than the solution $x$ to the equation $\mu([x, 1]) = 0.5$. These regulatory measures can be shaped by social policies, be determined by legislative entities or government bodies, and typically hinge on ensuring that the insurer's profit $\Pi$ remains non-negative for the regulation to be implementable.

From an economic perspective, the most natural regulatory approach involves finding $\theta$ and $R$ to maximize the social welfare. However, when the measure $\mu$ admits a density, we will observe later in Lemma 4 that the marginal cost is lower than the average cost. This situation, as explored in [26], leads to a deficit for the insurer. Consequently, the need arises to devise a second-best public pricing rule to address this concern, as discussed in [14], [35], and [43]. We adopt in this paper an innovative approach where the regulator first selects $R$, and subsequently, the insurer determines $\theta$. For a specific $R$, the insurer will optimize the single-variable function $\theta \mapsto \Pi(R,\theta)$ to find the value $\theta_R^*$ that maximizes the function. The regulator can anticipate this choice of $\theta_R$ and set $R$ accordingly to ensure that $\mathrm{SW}(R,\theta_R^*)$ attains its maximum value[5].

When $\mu$ admits a continuous density on $(0,1)$, the existence of an optimal contract is easy to prove in the two following regulatory regimes.

**Fact 4.** *Consider a probability measure $\mu$ that admits a continuous density $f$ on $(0,1)$. The following results hold:*

*(i) When the indemnity $R \in (0, L]$ is regulated, representing an exogenous variable for the insurer, the insurer maximizes its total profit by solving an uni dimensional optimization problem and thus seeks for the optimal market segmentation $\theta_R^*$. The optimal contract exists and is denoted $C_R^* = (\bar{P}(\theta_R^*, R), R)$.*

*(ii) When there is no regulation, the insurer maximizes its total profit by solving an optimization problem in two variables and seeks an optimal market segmentation and an optimal indemnity, that is, $(\theta^*, R^*)$. The optimal contract exists and is denoted $C^* = (\bar{P}(\theta^*, R^*), R^*)$.*

*In both scenarios, an optimal contract emerges from the profit-maximization pursued by the insurer.*

*Proof —* This result is a consequence of the Extreme Value Theorem. The continuity of $\Pi$ is a direct consequence of the continuity of $\bar{P}$ and the measure $\mu$, with boundedness being an inherent feature as well. For the unregulated scenario (case ii), the existence result follows from the compactness of the domain $[0,1] \times [0,L]$. In the regulated case (case i), the existence of an optimal contract is guaranteed for every $R \in [0,L]$, by the compacity of $[0,1]$ combined with the continuity and boundedness of the one-variable function $\theta \mapsto \Pi(R,\theta)$. □

It remains uncertain at this stage whether the optimal contracts $C^*$ or $C_R^*$ are profitable. In other words, it is not guaranteed that the profit derived from offering these optimal contract $C_R^*$ will be positive. In both regulatory and non-regulatory scenarios, the insurer would avoid generating a deficit by simply selecting $\theta^* = 1$ (or $\theta_R^* = 1$) as their optimal segmentation. In practical terms, it would be similar to the insurer opting not to provide an insurance contract. This situation calls for investigating profitability regions, which is what we shall do in the next sections.

## 4. Average probability of damage, cost and profit functions in the regular case

Throughout the subsequent sections of this paper, we will make the assumption that the probability measure $\mu$ is regular. Specifically, the associated distribution function is assumed to be continuous on $[0,1]$ and continuously differentiable on $(0,1)$. Later on, we shall denote by $\mathcal{M}_1$ this set of underlying probability measures. Note that $\mu([0,\theta]) = F(\theta) - F(0) = F(\theta)$ for every $\theta \in [0,1]$. Given the twice continuous differentiability and the increasing nature of $F$ over $[0,1]$, the density function $f$ is defined and positive on the interval $(0,1)$ and $f(\theta) = F'(\theta)$. It is easy to show that the profit function, as defined in equation (9)) can be written a

$$\Pi(\theta, R) = \underbrace{\left[\bar{P}(\theta, R) - R \times A(\theta)\right]}_{\text{Average profit}} \underbrace{[1 - F(\theta)]}_{\text{Quantity}} \tag{12}$$

---

[5]It should be pointed out that we make the implicit assumption that the insurer and the regulator share the same information, see [33].

where $A(\theta)$ is the average probability of the group of policy holders and $Q(\theta) = 1 - F(\theta)$ is the total quantity of contracts sold. Note that $A$ is differentiable and $Q'(\theta) = -f(\theta) < 0$, which implies that when $\theta$ increases, the quantity (of contract) decreases.

*4.1. The Average Probability of Damage and its Properties*

Since the average probability of damage will play a critical role in this paper, we provide a formal definition even though it is fairly well-known.

**Definition 2.** *Let A be defined on the interval $[0, 1)$ by*

$$A(\theta) = \frac{1}{1 - F(\theta)} \int_\theta^1 x f(x) \, dx \tag{13}$$

*For $\theta \in [0, 1)$, $A(\theta)$ is the average probability of damage of the group of policy-holders $[\theta, 1]$.*

From a pure probabilistic point of view, this average probability of damage, sometimes called a truncated mean (see [16]) or a left-truncated mean (see [2]), is a simple conditional expectation[6].

In our insurance framework, for any selected $\theta \in [0, 1)$, the group of policyholders $G = [\theta, 1]$ is *heterogeneous* due to the distinct probabilities of damage across agents. In practice, this heterogeneity prompts insurance companies to assemble individuals with comparable risk profiles into the same group. This practice involves pooling a substantial number of policyholders based on their observable attributes to mitigate fluctuations around the average group probability (for more practical insights, refer to [40]). When the count of agents is finite, as in real-world situations, the application of (exact) laws of large numbers is impossible, resulting in a deviation from the mean within the group. Yet, given our focus on a *continuum* of types, as long as the group encompasses an interval structure like $G = [\theta, 1]$ (excluding singleton cases), the cardinality of the group aligns with the *continuum*. In this context, an adapted version of the law of large numbers becomes applicable[7]. This implies that, informally speaking, the average probability within the group tends towards $A(\theta)$ with almost certain probability. Consequently, when a specific $R$ and $\theta$ are chosen, the insurer can confidently anticipate that the claims' cost will be $R \times A(\theta)$. All the quantities become deterministic. In that sense, there is no need to talk about expected cost of claims.

Thanks to the regularity of $F$, the function $A$ is twice continuously differentiable on $(0, 1)$. The derivative can be computed explicitly:

$$\begin{aligned} A'(\theta) &= \frac{f(\theta)}{(1 - F(\theta))^2} \int_\theta^1 x f(x) \, dx - \frac{\theta f(\theta)}{1 - F(\theta)} \\ &= (A(\theta) - \theta) \frac{f(\theta)}{1 - F(\theta)} \end{aligned} \tag{14}$$

**Definition 3.** *Let the mean residual lifetime function, denoted $m$, and the hazard rate or failure rate, denoted $H$, be defined on $[0, 1)$, as follows:*

$$m(\theta) = A(\theta) - \theta \quad \text{and} \quad H(\theta) = \frac{f(\theta)}{1 - F(\theta)} \tag{15}$$

---

[6]If $X$ is the underlying random variable whose distribution function if $F$, then, $A(\theta)$ is by definition equal to $\mathbb{E}(X | X \geq \theta)$, a conditional expectation. Using Bayes rule (assuming that $x > \theta$), it is easy to show that $F(x | X \geq \theta) = (F(x) - F(\theta))/(1 - F(\theta))$. Differentiating $F$ with respect to $x$ leads to the conditional density equal to $f(x | X \geq \theta) = f(x)/(1 - F(\theta))$. It suffices to integrate to find $A(\theta)$. Note that $A$ is a non-dimensional quantity.

[7]It is important to note, as highlighted in [34], that the application of laws of large numbers warrants careful consideration. The use of a strong law of large numbers presents challenges due to potential measurability issues. However, following the approach in [52], a weaker version of the law of large numbers ($\mathcal{L}_1$-convergence) can be established, as long as the group forms an interval rather than a singleton.

We refer to [3] (see also [2]) for a review of the relations between these quantities. With these definitions, we have:
$$A'(\theta) = H(\theta) \times m(\theta) \tag{16}$$
This is consistent with what can be found in the literature (see for instance [30], equation (2.4)).

At this point, it is worth noting the situation at the endpoints of the interval $[0,1]$. The cumulative distribution function $F$ is defined on $[0,1]$ and twice continuously differentiable on $(0,1)$, The functions $A$ is defined and is continuous on $[0,1)$ but not at $1$ where it is undefined since the denominator would be zero in (13) when $\theta = 1$. The functions $m$ and $H$ are also defined on $[0,1)$. However, we show in Lemma 7 in Appendix B, that $A$ has a finite limit equal to 1 as $\theta$ approaches 1. Consequently, we can extend $A$ by continuity to define it over the entire closed interval $[0,1]$. Specifically, $A$ is defined by (13) on $[0,1)$, and we set $A(1) = 1$. This extended definition holds economic significance: when only the agent $\theta = 1$ purchases the contract, the average probability naturally becomes 1.

Consequently, the average probability of damage $A$ and the mean residual lifetime function $m$ are now defined and continuous on $[0,1]$. It's important to note that the differentiability of $A$ (as well as $m$) is only ensured on the open interval $(0,1)$. The differentiability of $A$ at the endpoint $\theta = 1$ requires further assumptions on the density function $f$, that we explain in the following lemma. We present more details in Appendix B.

**Lemma 2.** *If the density function $f$ is equivalent, in 1, to a power function with exponent $s > -1$, then $A$ is differentiable in 1 and*
$$A'(1) = \frac{s+1}{s+2}.$$

The differentiability of $A$ at 1 and the value $A'(1)$ will play a central role in the subsequent section as we explore market profitability. Additionally, this differentiability will prove valuable during numerical computations. Particularly, when $\theta$ is close to 1, both $1 - F(\theta)$ and $\int_\theta^1 x f(x)\, dx$ in (13) become very small, potentially leading to poor numerical outcomes. In [11], we address this by approximating $A$ using its first-order Taylor expansion for values of $\theta$ near 1.

To illustrate Lemma 2, we can provide an example using the Beta distribution with parameters $\alpha > 0$ and $\beta > 0$. This is done in Corollary 2 in Appendix B, where we obtain
$$A'(1) = \frac{\beta}{\beta+1} \tag{17}$$

**Remark 2.** *In the last section devoted to the numerical analysis, we shall assume that the types are distributed according to a two-parameter Beta distribution. Instead of working with the natural parameters $(\alpha, \beta) \in \mathbb{R}_+^2$, we shall make a change of variable in order for the set of parameters to lie in a bounded subset of $\mathbb{R}_+^2$, indeed the unit square. The change of variable will make use of $A'(1) = \frac{\beta}{\beta+1}$.*

The differentiability of $A$ at 0 requires further investigation, which is discussed in Appendix B. While this aspect may be of lesser significance for the present study, it is still interesting to address.

**Lemma 3.** *For all $\theta \in (0,1)$*

(i) $m(\theta) > 0$ and $A(\theta) > \theta$

(ii) $A'(\theta) > 0$

(iii) $H(\theta) > 0$

(iv) *If $A''(\theta) \geq 0$ then $H'(\theta) > 0$, but the converse is not true.*

(*i*) and (*iv*) are proven in the appendix. (*ii*) and (*iii*) are simple observations of equations (16) and (15). Note the hypothesis in (*iv*), impose the average probability $A$ to be linear or convex. Let us consider three levels of hypothesis for our probability measure of the risk of the agents.

**Definition 4.** *Let*

- $\mathcal{M}_1$ *be the set of probability measures with a cumulative distribution function $F$ that is continuous on the interval $[0,1]$ and twice continuously differentiable on the open interval $(0,1)$.*

- $\mathcal{M}_2$ *be the set of probability measures in $\mathcal{M}_1$ whose density function $f$ is equivalent, in 1, to a power function with exponent $s > -1$,*

- $\mathcal{M}_3$ *be the set of probability measures in $\mathcal{M}_2$ whose associated average probability of damage $A$ satisfies $A''$ is non-negative on $(0,1)$.*

Most of the results in this paper are true for probability measures in $\mathcal{M}_2$. In parts of this paper, we wish to be in the situation where the hypotheses of Lemmas 2 and 3 (iv) are met, and we will require the probability measure to be in $\mathcal{M}_3$.

It should be pointed out that $\mathcal{M}_3 \subset \mathcal{M}_2 \subset \mathcal{M}_1$ and $\mathcal{M}_3$ is not empty. For instance, the continuous uniform distribution satisfies all three conditions mentioned above. Furthermore, in Lemma 11, we demonstrate that all Beta distributions with $\alpha = 1$ and $\beta > 0$ belong to $\mathcal{M}_3$. We also have forthcoming Conjecture 1, supported by numerical verification [11], that all Beta distributions with $\alpha > 1$ are part of $\mathcal{M}_3$, although we don't have a formal proof. In addition to these distributions, the set $\mathcal{M}_3$ includes a variety of probability measures.

*4.2. Cost Functions and the Natural Monopoly Problem*

For a given $\theta \in [0,1]$ and $R \in [0,L]$, let $\text{TC}(\theta, R)$ be the total cost (see (9)), equal to

$$\text{TC}(\theta, R) = R \int_\theta^1 x f(x), dx = R \cdot A(\theta) \cdot Q(\theta) \tag{18}$$

where $Q(\theta) = 1 - F(\theta)$ represents the total quantity of contracts sold to the group of agents $[\theta, 1]$. By definition, the average cost denoted $\text{AC}(\theta, R)$ is the total cost divided by the quantity:

$$\text{AC}(\theta, R) = R \cdot A(\theta) \tag{19}$$

The derivation of the marginal cost is slightly trickier and we relegate the proof in Appendix A.

**Lemma 4.** *The marginal cost is equal to*

$$MC(\theta, R) = \theta R \tag{20}$$

*and we further have*

$$AC(\theta, R) > MC(\theta, R)$$

The fact that the average cost is always greater than the marginal cost indicates that the insurer is in a natural monopoly situation. Consequently, setting the premium equal to the marginal cost would inevitably lead to a financial deficit. This natural monopoly situation thus gives rise to a second best pricing problem as discussed in works such as [14] and [43]. It's worth noting that the concept of a natural monopoly, although not explicitly labeled as such, has been recognized in the insurance field in studies like [26] and [22].

In [26], Figure 2 illustrates a scenario where both the marginal cost and average cost are assumed to be linear functions of quantity. This specific instance corresponds to our model when the density function $f$ represents a uniform distribution, although this assumption is not the most realistic. To elaborate, when $f$ is a uniform density on the interval $(0, 1)$, the cumulative distribution function $F$ simplifies to $F(\theta) = \theta$, resulting in $Q(\theta) = 1 - \theta = q$, where $q$ denotes the quantity corresponding to a given $\theta$. Consequently, the marginal cost as a function of $q$ is given by $\overline{\mathrm{MC}}(q, R) = R(1-q)$, and it is easy to show that $\overline{\mathrm{AC}}(q, R) = R(1 - q/2)$. In both cases, the cost functions exhibit linearity.

*4.3. Average Profit Function $\mathcal{R}$*

At this stage, we only assume that $\mu \in \mathcal{M}_1$, that is, $F$ is continuous on the interval $[0, 1]$ and twice continuously differentiable on the open interval $(0, 1)$. In this case, the profit function $\Pi$ is in $\mathcal{C}^2((0,1) \times (0, L])$. It's worth noting that while the differentiability of $A$ at 1 is not a requirement here, having it ensures that $\Pi$ is in $\mathcal{C}^2((0,1] \times (0, L])$. Let us note $\mathcal{R}$ the *average profit function* appearing in (12):

$$\mathcal{R}(\theta, R) = \bar{P}(\theta, R) - R \times A(\theta) \tag{21}$$

With this, we can express the profit function as:

$$\Pi(\theta, R) = \mathcal{R}(\theta, R) \times (1 - F(\theta)) = [\bar{P}(\theta, R) - R \times A(\theta)](1 - F(\theta)) \tag{22}$$

The function $\mathcal{R}$ will prove important in the analysis of the insurance market's profitability.

**Definition 5.** *The single-contract insurance market is profitable if there exists an admissible contract $C = (\theta, R)$ such that $\Pi(\theta, R) > 0$.*

In a framework similar to ours, initially formulated in [46], [32] establishes a no-trade condition that leads to the conclusion that the endowment $\{(W_0 - L, W_0)\}$ is the only implementable allocation, indicating that no profitable insurance market can exist. In our analysis, we define a notion of profitability that is more constrained than the no-trade condition presented in [32], primarily because we are considering a single contract. It's important to clarify that the lack of profitability for a single contract doesn't inherently eliminate the possibility of a profitable set of contracts (even if not countable), as discussed in works such as [46] and [18]. Notably, as it is well known, in cases where agents are risk-neutral, an insurance market can't be profitable. This is stated in the following fact, which is proven in Appendix B.

**Fact 5.** *When agents are risk-neutral, regardless of the density $f$, there is no admissible contract $C = (P, R)$ for which $\Pi(P, R) > 0$.*

For the insurance problem to be relevant, agents must be risk-averse but risk aversion is in general not a sufficient condition. Interestingly, we shall show that a sufficient condition for the profitability of the insurance market is related to $A'(1)$ when $A$ is differentiable in 1. We shall assume $\mu$ leads to $A$ being differentiable in 1 (for instance it belongs to $\mathcal{M}_2$) in the rest of this section.

## 5. The Optimal Insurance Premium for a Given Indemnity as an Inverse Elasticity Rule

The indemnity $R \leq L$ is here assumed to be given, which means that only the segmentation variable $\theta$ is chosen.

*5.1. Profitability analysis*

Noting that for each $\theta \in [0, 1)$, $Q(\theta) > 0$ while $Q(1) = 0$, it is easy to see from equation (22) that the following equivalences are true:

$$\forall \theta \in [0, 1], \quad \Pi(\theta, R) \geq 0 \iff \mathcal{R}(\theta, R) \geq 0$$

$$\forall \theta \in [0, 1), \quad \Pi(\theta, R) > 0 \iff \mathcal{R}(\theta, R) > 0$$

From part *(i)* of Proposition 1, we have $\mathcal{R}(1, R) = \bar{P}(1, R) - RA(1) = 0$, that is, the market segment is reduced to the singleton $\{1\}$ and the profit of the insurer is zero. The interesting question is whether there exists $\theta \in (0, 1)$ for which the insurer makes a (strictly) positive profit. As explained above, this is equivalent to having $\mathcal{R}(\theta, R) > 0$. Define:

$$\hat{\Theta}(R) = \{\theta \in [0, 1) : \mathcal{R}(\theta, R) > 0\} \tag{23}$$

This set is the preimage of the open set $(0, +\infty)$ by the continuous function $\mathcal{R}$. Therefore it is an open set. As such, it can be expressed as a union of disjoint open intervals. The separability property of the real line guarantees that the count of such intervals is either finite or countably infinite, although this information doesn't have a direct impact on the subsequent discussion. Here are the two possible scenarios to consider regarding the region of profitability, i.e. the set $\hat{\Theta}(R)$:

*(i)* The set $\hat{\Theta}(R)$ is empty: In this situation, the market is unprofitable. An illustrative example of this scenario can be observed in Figure 2 (bottom left).

*(ii)* The set $\hat{\Theta}(R)$ consists of one interval or more: In this situation, the market is profitable.

- If the upper bound of the interval is equal to 1, we shall denote $\hat{\theta}_R$ its lower bound, so the region of profitability is $(\hat{\theta}_R, 1)$. This situation is visually represented in Figure 2 (top left) and (bottom right).

- Alternatively, no interval has 1 has an upper bound. In this context, the insurer identifies a market segment $\theta$ within this interval, and all agents falling within the range $[\theta, 1]$ will choose to purchase the insurance. This scenario is illustrated in Figure 2 (top right)

This discussion and the examination of Figure 2, suggests that investigating the behavior at 1 can yield a viable condition for establishing profitability. The following proposition, established in Appendix B, offers a sufficient condition to ensure profitability. This condition is base on the way two curves meet at the point 1: the blue curve representing the critical price $\bar{P}(\theta, R)$ and the red curve representing the average risk $A(\theta)$.

**Proposition 2.** *If $A$ is differentiable in 1 and*

$$RA'(1) > \frac{\partial \bar{P}}{\partial \theta}(1, R) \tag{24}$$

*There exists $\hat{\theta}_R \in (0, 1)$ such that $(\hat{\theta}_R, 1)$ is profitable.*

Note the proposition indicates $(\hat{\theta}_R, 1)$ is one of the components of $\hat{\Theta}(R)$, thereby establishing that $\hat{\Theta}(R)$ is non-empty. However, we cannot assert or dismiss the presence of additional intervals within $\hat{\Theta}(R)$.

To illustrate this proposition, let us consider the case of a CARA utility. The expression for the willingness to pay, denoted as $\bar{P}$, is provided by Equation (8). Let $\Xi_R(\lambda) = \frac{1}{R} \frac{\partial \bar{P}}{\partial \theta}(1, R)$. It is not difficult to show that

$$\Xi_R(\lambda) = \frac{e^{-\lambda L}(1 - e^{-\lambda R})}{\lambda R} \tag{25}$$

Equation (24) thus becomes $A'(1) > \Xi_R(\lambda)$. Since $\Xi_R$ is monotonically decreasing and $\lim_{\lambda \to \infty} \Xi_R(\lambda) = 0$, there exists a certain minimum value of $\lambda$ for which the profitability condition is satisfied. Conversely, $\lim_{\lambda \to 0} \Xi_R(\lambda) = 1$, therefore the profitability condition simplifies to $A'(1) > 1$, and thus it can never be met for a $\lambda$, which is small enough. This situation is succinctly encapsulated in the following fact:

**Fact 6.** *Assume a CARA utility so that the willingness to pay is given in equation (8). The profitability condition given in equation (24) is always met when $\lambda$ is high enough and is never met when $\lambda$ is low enough.*

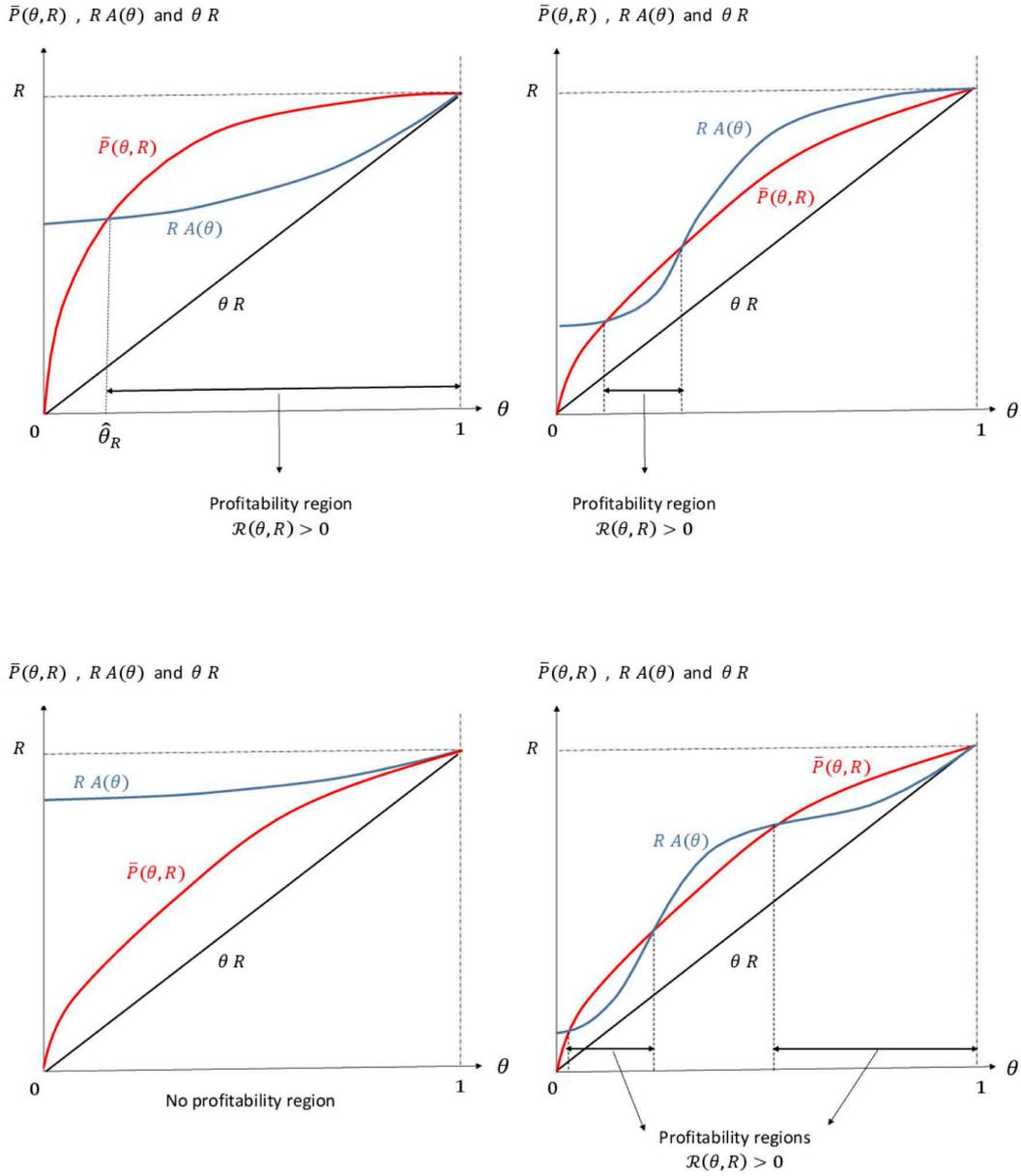

Figure 2: (Top-Left) $A$ is convex and there is a profitable region $[\widehat{\theta_R}, 1)$   (Top-Right) $A$ is not convex and there is a profitable region whose right endpoint is not 1   (Bottom-Left) $A$ is not convex and there is no profitable region   (Bottom-Left) $A$ is not convex and there the profitable region is not convex.

*5.2. Optimal Market Segmentation and Hazard Rate*

We know the profit function $\Pi$ is continuous with respect to $\theta$. Given that the interval $[0, 1]$ is a compact set, the extreme value theorem guarantees that $\theta \mapsto \Pi(\theta, R)$ has a maximum for each specific $R$. In instances where $\hat{\Theta}(R)$ is not empty (indicating market profitability), an optimal segmentation denoted as $\theta_R^*$ exists within $\hat{\Theta}(R)$. This segmentation satisfies the condition that $\theta \mapsto \Pi(\theta, R)$ attains its positive maximum at $\theta_R^*$, signifying $\Pi(\theta_R^*, R) > 0$. In the subsequent result, proven in Appendix B, we offer two distinctive characterizations of the optimal profitable segmentation $\theta_R^* \in \hat{\Theta}(R)$, which corresponds to the first-order condition. Here $H(\theta)$ represents the hazard rate assessed at $\theta$, as defined in (15).

**Lemma 5.** *Assume that $\hat{\Theta}(R) \neq \emptyset$. The two following statements offer a necessary condition of optimal segmentation $\theta_R^* \in \hat{\Theta}(R)$:*

$$\frac{\frac{\partial \bar{P}}{\partial \theta}(\theta_R^*, R) - R\,A'(\theta_R^*)}{\bar{P}(\theta_R^*, R) - R\,A(\theta_R^*)} = H(\theta_R^*) \tag{26a}$$

$$\frac{\frac{\partial \bar{P}}{\partial \theta}(\theta_R^*, R)}{\bar{P}(\theta_R^*, R) - R\,\theta_R^*} = H(\theta_R^*) \tag{26b}$$

At this stage, it is only assumed that $\hat{\Theta}(R) \neq \emptyset$ so that $\theta_R^*$ may not be unique. One reason may be that $\hat{\Theta}(R)$ is the union of more than one disjoint intervals or simply that several $\theta$ satisfy (26). As we will see, the uniqueness of the optimal of segmentation is guaranteed by the convexity (or linearity) of the average probability of damage $A$ and the concavity of the willingness to pay, denoted as $\bar{P}$.

We have established in Lemma 4 that the average cost is always greater than the marginal cost. Since $\theta_R^* \in \hat{\Theta}(R)$, the total profit is positive, and this means that both denominators appearing in equations (26) must be positive. Since the average cost and the marginal cost evaluated in $\theta_R^*$ are respectively equal to $R\,A(\theta_R^*)$ and to $R\,\theta_R^*$, the following corollary is true.

**Corollary 1.** $\bar{P}(\theta_R^*, R) > \text{AC}(\theta_R^*) > \text{MC}(\theta_R^*)$

From Lemma 5, an optimal market segmentation $\theta_R^*$ is such that the left hand side of equations (26) are equal to the hazard rate $H$, evaluated in $\theta_R^*$. While the hazard rate frequently appears in the first order condition of an optimization problem (see e.g., [18]), its economic interpretation remains unclear.

*5.3. Elasticity of Demand and the Unique Optimal Insurance Premium*

In economics textbooks, the monopolist is presumed to know the demand function denoted as $P \mapsto Q(P)$, where $P$ is the price chosen by the monopolist, thus exogenous, and $Q$ the resulting quantity sold, thus the endogenous variable. When the price increases $P$ to $P + h$, where $h \neq 0$, the elasticity is expressed as

$$\epsilon(P) = \frac{\Delta Q}{h} \times \frac{P}{Q(P)}$$

where $\Delta Q = Q(P + h) - Q(P)$. In cases where the demand function $Q$ is continuously differentiable ($\mathcal{C}^1$), as the increment $h$ approaches zero, we obtain the well-known elasticity of demand defined as

$$\epsilon(P) = Q'(P) \times \frac{P}{Q(P)}.$$

In our framework, the exogenous variable is not $P$ but $\theta_R$. Thus, the elasticity is defined as a function of $\theta_R$. It is the the ratio of relative variations as we shift from $\theta_R$ to $\theta_R + h$:

$$\frac{\frac{\Delta Q(.)}{Q(.)}}{\frac{\Delta \bar{P}(.)}{\bar{P}(.)}} = \frac{\frac{Q(\theta_R + h) - Q(\theta_R)}{Q(\theta_R)}}{\frac{\bar{P}(\theta_R + h, R) - \bar{P}(\theta_R)}{\bar{P}(\theta_R, R)}}$$

Dividing $Q(\theta_R + h) - Q(\theta_R)$ and $\bar{P}(\theta_R + h, R) - \bar{P}(\theta_R)$ by $h$ and letting $h$ goes to zero, using the fact that the functions $Q$ and $\bar{P}$ are $\mathcal{C}^2$ on $(0,1)$, the elasticity evaluated in $\theta_R$ is equal to

$$\epsilon(\theta_R) = \frac{Q'(\theta_R)}{\frac{\partial \bar{P}}{\partial \theta}(\theta_R, R)} \times \frac{\bar{P}(\theta_R, R)}{Q(\theta_R)} < 0 \qquad (27)$$

which is negative since $Q'(\theta_R) = -f(\theta_R) < 0$ while all the other quantities involved are positive.

Our framework is more complex than the classical monopoly theory since the only exogenous variable chosen by the insurer is the market segmentation $\theta_R$. The quantity and the premium are functions of $\theta_R$ and this means that both the quantity and the premium are *endogenous variables*. This explains the presence of the additional (derivative) term $\frac{\partial \bar{P}}{\partial \theta}(\theta_R)$ in the elasticity formula. Note that the slope of the demand function is equal to $Q'(\theta_R)/\frac{\partial \bar{P}}{\partial \theta}(\theta_R, R)$ and is negative, as expected. It is easy to show that equation (27) is equivalent to

$$\epsilon(\theta_R) = -H(\theta_R) \times \frac{\bar{P}(\theta_R, R)}{\frac{\partial \bar{P}}{\partial \theta}(\theta_R, R)} = -H(\theta_R) \times \frac{1}{\frac{\partial}{\partial \theta}\left[\ln(\bar{P}(\theta_R, R))\right]} < 0 \qquad (28)$$

We believe this equation presents an interesting formulation of the elasticity because it provides a decomposition of elasticity into two distinct components: one tied to the distribution of types and the other linked to preferences, specifically risk aversion:

- The first term, $H(\theta_R)$, is the hazard rate and is associated with the "statistical aspect" of the market, reliant solely on the distribution of types.

- The second term, expressed as $1/\frac{\partial}{\partial \theta}\left[\ln(\bar{P}(\theta_R, R))\right]$ captures the the "preference aspect" of the market. This term relies only on willingness-to-pay, which is contingent upon the utility function $U$, thus reflecting the underlying level of risk aversion of the agents.

With this elasticity, we establish in the subsequent proposition proven in Appendix B, a set of sufficient conditions to ensure the uniqueness of optimal segmentation. It is important to point out that, as in classical monopoly theory, the optimal premium satisfies the Lerner index, which is an inverse elasticity rule. Recall the marginal cost is defined as $MC(\theta, R) = \theta \times R$.

**Proposition 3.** *Assume $U''$ is increasing, $\mu \in \mathcal{M}_3$ and $R A'(1) > \frac{\partial \bar{P}}{\partial \theta}(1, R)$, then*

- *there exists a unique $\hat{\theta}_R \in (0,1)$ such that $\hat{\Theta}(R) = (\hat{\theta}_R, 1)$,*
- *there exists a unique optimal profitable segmentation $\theta_R^* \in (\hat{\theta}_R, 1)$,*

*such that the optimal premium charged $\bar{P}(\theta_R^*, R)$ satisfies the inverse elasticity rule:*

$$\bar{P}(\theta_R^*, R) \times \left(1 + \frac{1}{\epsilon(\theta_R^*)}\right) = \text{MC}(\theta_R^*, R) \qquad (29)$$

*and we further have $\epsilon(\theta_R^*) < -1$ (or $|\epsilon(\theta_R^*)| > 1$).*

To the best of our knowledge, the formulation of the optimal premium as the classical Lerner index (inverse elasticity rule) within an insurance monopoly framework is new and unprecedented. Notably, we are able establish that the optimal premium is unique and satisfies the inverse elasticity rule even though the premium remains implicitly defined in general. In a paper devoted to optimal contract regulation within a competitive framework, [56] show that when the density $f$ is log-concave, then the equilibrium is unique. Within our model, we don't assume that the density $f$ is log-concave. We assume instead that $\mu \in \mathcal{M}_3$, i.e., $f$ is such that $A''$ is non negative on $(0,1)$. Under this assumption, we know from lemma 3, $(iv)$ that the hazard rate is increasing and we show that the average profit function $\mathcal{R}$ is log-concave (with respect to $\theta_R$). As a result, the optimal premium is unique.

When the insurer offers a single insurance contract, the optimal premium is greater than the marginal cost, as in the classical theory of monopoly. Since it is a natural monopoly situation, the premium must also be greater than the average cost for the insurer to make positive profits. From equation (29), it is easy to see that

$$\theta_R^* = \frac{\bar{P}(\theta_R^*, R)}{R} \times \left(1 + \frac{1}{\epsilon(\theta_R^*)}\right) \tag{30}$$

As one may intuitively expect, in equation (30), the elasticity directly impacts the (optimal) premium-to-coverage ratio $\bar{P}(\theta_R^*, R)/R$.

## 6. Choice of the Indemnity: Profit versus Social Welfare Maximization

In this section, we continue to examine the scenario where the insurer is obligated to provide a single contract. However, if community rating is subject to additional regulatory measures, the selection of the indemnity might rest with the regulator rather than the insurer. Naturally, in the absence of further regulatory restrictions on community rating, the monopolistic insurer retains the autonomy to determine the indemnity along with the segmentation, making the choice with the intention of profit maximization.

### 6.1. Profit Maximization

In the absence of additional regulatory limitations on community rating, the insurer is constrained to designing only a single contract but has the liberty to independently select the indemnity $R \in (0, L]$. As previously mentioned, the insurer's objective will be to maximize the total profit function, defined as $(\theta, R) \mapsto \Pi(\theta, R)$ (refer to equation 22).

Let

$$\hat{\Theta} := \bigcup_{R \in (0, L]} \hat{\Theta}(R) \tag{31}$$

be the overall profitability region where $\hat{\Theta}(R)$ is defined in (23). as the collective profitability region, where $\hat{\Theta}(R)$ is as defined in (23). It's worth noting that although the union in equation (31) is not necessarily countable, $\hat{\Theta}$ is equivalently represented as the projection onto the first axis:

$$\hat{\Theta} = \{(\theta, R) \in [0, 1) \times [0, L] : \mathcal{R}(\theta, R) > 0\} \tag{32}$$

This set is open because it is the preimage of the open set $(0, +\infty)$ by the continuous function $\mathcal{R}$, therefore the definition provided in (31) doesn't raise any concerns regarding the measurability of $\hat{\Theta}$. Importantly, whenever there exists an $R \in (0, L]$ for which $\hat{\Theta}(R) \neq \emptyset$, it follows that $\hat{\Theta}$ itself is non-empty.

In the next lemma, we derive the properties of the optimal contract when the indemnity is not regulated.

**Lemma 6.** *If $(\theta^*, R^*) \in \hat{\Theta}$ maximizes the total profit function given in equation (22), then*

$$\bar{P}(\theta^*, R^*) \times \left(1 + \frac{1}{\epsilon(\theta^*)}\right) = \underbrace{\mathrm{MC}(\theta^*, R^*)}_{\theta^* \ R^*} \tag{33a}$$

$$\frac{\partial \bar{P}(\theta^*, R^*)}{\partial R} = A(\theta^*) \tag{33b}$$

*Proof —* The first order optimality conditions require that both partial derivatives $\frac{\partial \Pi}{\partial \theta}$ and $\frac{\partial \Pi}{\partial R}$ vanish at a potential optimal point $(\theta^*, R^*) \in \hat{\Theta}$, should such a point exist. Solving for $\frac{\partial \Pi}{\partial \theta}(\theta^*, R^*) = 0$ leads to (33a), employing arguments and computations analogous to those applied in Proposition 3 for a given $R$. Regarding $\frac{\partial \Pi}{\partial R}(\theta^*, R^*) = 0$, we use equation (22). It is straightforward to establish that, for each $\theta \in (0, 1)$, the condition $\frac{\partial \Pi(\theta, R)}{\partial R} = 0$ is equivalent to (33b). □

When the indemnity is not regulated, Lemma 6 provides a simple overall picture of the insurer's optimal choice. The optimal market segmentation $\theta^*$ and the optimal indemnity $R^*$ should be chosen such that

- The premium satisfies the so-called Lerner index, i.e., the inverse elasticity rule.
- The derivative with respect to $R$ of the willingness-to-pay should be equal to the average probability of the group of policy-holders.

Recall that in Proposition 3, we have exhibited conditions under which the optimal market segmentation $\theta_R^*$ is unique. In lemma 6, no conditions have been offered for the uniqueness of $R^*$.

*6.2. Social Welfare Maximization*

We will assume that the regulator possesses the same information to that of the insurer, [33, p. 76], which means that the regulator is in a position to replicate the computation of the private insurer.

For an indemnity $R \in [0,L]$ and a market segmentation $\theta \in (0,1)$ (thus for a given contract $C = (\bar{P}(\theta,R), R)$, the surplus SP$(\theta,R)$ of the policy holders is given by (10) and the social welfare SW$(\theta,R)$ is given by (11). As usual in Economics when one considers a regulated firm (e.g., [14], [35], [43]), it is natural for the regulator to maximize the social welfare, that is

$$\max_{(\theta,R) \in [0,1] \times [0,L]} \text{SW}(\theta, R) \qquad (34)$$

This maximization problem has a solution since SW is a continuous function and $[0,1] \times [0,L]$ is a compact set. An optimum $(\theta^{**}, R^{**})$ from the social point of view will satisfy that both partial derivative of SW vanish in this point. A necessary condition is $\partial SW(\theta^{**}, R^{**})/\partial \theta = 0$ given by

$$-\bar{P}(\theta^{**}, R^{**})f(\theta^{**}) + R^{**}\theta^{**}f(\theta^{**}) = 0$$

Subsequently, a regulation derived from social welfare maximization thus will implement the so-called marginal cost pricing rule [43]

$$\bar{P}(\theta^{**}, R^{**}) = \text{MC}(\theta^{**}, R^{**}) \qquad (35)$$

As we have mentioned already, since the insurer is in a natural monopoly situation in that the average cost is always greater than the marginal cost, the marginal cost pricing would generate a deficit, and this means that this first best (marginal cost) pricing is not feasible. One must thus look for a second best pricing. In the literature on the subject, it is usual to consider the problem as defined in equation (34) but subject to a profitability constraint, that is, the (insurer) profit should be greater than a positive constant (see classical textbooks on the subject, [35], [43], [14]). We shall in what follows take a different road map. We shall assume that $\theta_R^*$ (unique under some conditions, see Proposition 3) maximizes the total profit but the indemnity $R$ is then chosen to maximize the social welfare by the regulator.

$$\max_{R \in \hat{\Theta}} \text{SW}(\theta_R^*, R) \qquad (36)$$

If this problem has a solution, the resulting profit will necessarily be positive. It is however difficult to explicitly solve this optimization problem given in equation (36) since it remains unclear whether or not $R \mapsto \theta_R^*$ can be function.

Simplification arises when the analysis is confined to a CARA utility function, which is the specific scenario we will focus on for the numerical analysis. According to Fact 6, it's established that in the context of CARA utility, the contract is invariably profitable as long as the value of $\lambda$ is sufficiently high.

**7. Comprehensive Numerical Analysis with a Dimensionless Problem**

Up until now, most of the quantities have been treated as dimensional variables; that is, they have had specific units of measurement. For instance, $W$ is measured in a currency such as euros or dollars.

However, during the numerical analysis, we will work with non-dimensional variables. Employing such non-dimensional variables is a standard practice in engineering that offers several benefits. It allows for more accurate computations when some parameters are very small and prevents potential confusion and misleading interpretations arising from the interplay between dimensional and dimensionless variables.

For our numerical simulations, we will need to choose a utility function. We have opted for the Constant Absolute Risk Aversion (CARA) function. This function is characterized by a single positive parameter that denotes the level of risk aversion.

We also need to choose a distribution of probability for the agent type. Since the type $\theta \in [0,1]$ follows a continuous density $f$, the two-parameter Beta density emerges as a natural choice to conduct the numerical analysis. This choice arises from the density's support being $[0,1]$ and its remarkable flexibility that accommodates a diverse array of shapes. However, for a Beta distribution, the space of parameters is in $\mathbb{R}_+^2$. Since the intent is to conduct a *comprehensive* numerical analysis, we will need to map $\mathbb{R}_+^2$ to a bounded domain, as explained in [12]. As opposed to some cases, where the authors restrict their analysis $\mathbb{R}_+^2$ to $[0,4] \times [0,4]$ (see e.g., [25]), which is not very satisfactory and certainly not comprehensive. Following [12], we shall thus make a change of variable through a (differentiable) bijective mapping in order to work with bounded subset of $\mathbb{R}_+^2$, indeed $[0,1] \times [0,1]$. The change of variable will explicitly makes use of $A'(1)$ since we now have provided conditions under which $A'(1)$, which plays an important role and which is well-defined for a Beta distribution (see (17)).

*7.1. Preliminaries: Nondimensionalization of the Problem and Comprehensive Analysis*

*CARA Utility and Nondimensional Problem.* We make the assumption that the Bernoulli utility function $W \mapsto U(W)$ is a Constant Absolute Risk Aversion (CARA) function, that is, for a any positive wealth $W \geq 0$, it is equal to $U(W) = 1 - \exp(-\lambda W)$ where $\lambda > 0$ reflects the degree of risk aversion of the agent (called absolute risk-aversion coefficient) and is equal to $-U''(W)/U'(W)$ which is invariant with respect to $W$ for a CARA utility function. Considering that the dimensional variable $W$ is quantified in currency, $\lambda$ itself is also dimensional. Its unit should correspond to the *inverse* of the currency unit used, ensuring that $\lambda W$ is dimensionless[8]. We now explain the non-dimensionalization process.

Recall that $W_0 > 0$ is the initial wealth endowment of each agent and let us write the dimensionless quantity $\lambda W$, which is the product of two dimensional quantities $\lambda$ and $W$ as

$$\lambda W = \left(\frac{W}{W_0}\right) \times (\lambda W_0).$$

By defining $w = W/W_0 > 0$ and $\rho = \lambda W_0 > 0$, then, regardless of the initial wealth $W_0 > 0$, the following equality is true

$$w\rho = \lambda W \tag{37}$$

The interesting aspect of this elementary transformation is that now, both $w$ and $\rho$ are *dimensionless*. In the economic literature, although in general not presented as such but see [37], $\lambda$ is known as the Absolute Risk Aversion Coefficient and is a dimensional quantity while $\rho$ is known as the Relative Risk Aversion Coefficient (henceforth called RRA) and is a dimensionless quantity. Equipped with (37), we are now in a position to define the utility function $u$ whose argument is the dimensionless wealth $w$. It is the counterpart of $U$ whose argument is dimensional wealth $W$. The relation between the two utility functions should satisfy $U(W) = c\,u(w)$, where $c > 0$ is a constant. We will chose $c = 1/(1 - e^{-\rho})$ and we shall define the non-dimensionalized CARA utility by

$$u(w) = \frac{1 - e^{-\rho w}}{1 - e^{-\rho}}, \tag{38}$$

---

[8] As it is widely recognized, applying the exponential function (and more generally transcendental functions) to a dimensional quantity lacks logical meaning due to the absence of a defined unit of measurement. Considering the power series representation of $e^x$, namely $\sum_{n \in \mathbb{N}} x^n/n!$, this signifies that the addition of $x$ (in the unit of $x$) to $x^2$ (in the square of the unit of $x$), and onwards, is not feasible in terms of dimensional consistency. The unit of measurement of $\sum_{n \in \mathbb{N}} x^n$ and, consequently, of $e^x$ remains undefined.

For all $w$, the ratio $-u''(w)/u'(w)$ is constant and equal to $\rho$, the relative risk-aversion coefficient, a dimensionless quantity. Any affine transformation of $u$ would yields the same result. The rest of the non-dimensionalization process can be found in Appendix D. The non-dimensional variables will now be denoted $l, r, \bar{p}$

*Comprehensive Analysis in the EZ-square.* For the numerical analysis, we will model the distribution of types using a two-parameter Beta distribution. This choice is motivated by the wide range of shapes that can be represented with this distribution. As is well-known, a beta distribution can exhibit various shapes such as single-peaked (∩-shaped), increasing, decreasing, or even ∪-shaped, depending on the values of its parameters. The Beta distribution on the interval $[0,1]$ is parametrized by two coefficients $\alpha$ and $\beta$, both in the $(0, \infty)$ range, and the probability density function $f$ is given by:

$$f(\theta) = \frac{1}{B(\alpha, \beta)} \theta^{\alpha-1}(1-\theta)^{\beta-1} \qquad \theta \in [0, 1] \tag{39}$$

where $B$ is the Beta function used to compute the normalization constant $B(\alpha, \beta)$. Let $\zeta$ the underlying random variable with a Beta density $f$ with parameters $\alpha$ and $\beta$. While $\alpha$ and $\beta$ appear explicitly in the definition of $f$ and are the standard parameters to define the Beta distribution, they do not carry any particular meaning in the context of our insurance problem. In particular, when both $\alpha$ and $\beta$ vary, the mean, the mode and the variance of the underlying random variable $\zeta$ will also change. Moreover, the set of parameters is $\mathbb{R}_+^2$ and thus is unbounded. For these reasons, following [12], we shall introduce another space of parameters, called the *EZ*-square which is simply equal to $(0,1) \times (0,1)$. To construct the *EZ*-square, consider the mapping

$$\varphi : (\alpha, \beta) \rightarrow \left(\frac{\alpha}{\alpha+\beta}, \frac{\beta}{\beta+1}\right) \tag{40}$$

and define:

$$E := \frac{\alpha}{\alpha+\beta} = A(0) \in (0,1) \qquad \text{and} \qquad Z := A'(1) = \frac{\beta}{\beta+1} \in (0,1). \tag{41}$$

Simple arithmetic shows that the inversion of $\varphi$ (i.e., solving $(\alpha, \beta) = \varphi^{-1}(E, Z)$) yields:

$$\alpha = \frac{E \times Z}{(1-E)(1-Z)} \qquad \text{and} \qquad \beta = \frac{Z}{1-Z} \tag{42}$$

and defines a *unique element* $(\alpha, \beta) \in (0, +\infty) \times (0, +\infty)$. From an economic point of view, $A(0) = E$ represents the expected cost of claims when the entire set of agents $[0,1]$ is insured while $A'(1)$ is related to the variance of the underlying Beta distributed random variable $\zeta$. When one uses the initial set of parameters, the variance is equal to $\alpha\beta/[(\alpha+\beta)^2(\alpha+\beta+1)]$ while one uses the parameters $E$ and $Z$, the variance $\mathcal{V}$ is equal to

$$\mathcal{V} = \text{Var}(E, Z) = \frac{E(1-E)^2(1-Z)}{1-E(1-Z)} \tag{43}$$

By construction of the bijective (differentiable) mapping $\varphi$, the set of parameters of a given Beta distribution are now located in $(0,1) \times (0,1)$ called the *EZ*-square. As shown in Fig. C.9, the *EZ*-square can be decomposed in four regions delimited by the second diagonal $]\Omega_3, \Omega_6[$ ($Z = 1-E$) and the horizontal segment $]\Omega_1, \Omega_4[$ ($Z = \frac{1}{2}$) depending upon the shape of the density. We refer the reader for more to Appendix C.

*7.2. Stylized Facts on Risk Aversion and Loss Distribution*

*Risk aversion.* More than thirty years ago, [49] examined whether the RRA is constant or not using property liability insurance data of 15 countries. His results show that for most countries, the RRA lies between one and three at the exception of three countries, Canada, Italy and Sweden. In [38], the author offers a valuable table that provides a review of the estimation of the RRA found in the literature. Depending upon the types of data used (insurance, finance, survey...), the estimation of the RRA are very different. For instance, using cross-country series on insurance, in [50], the authors consider the case of 29 countries and find that the RRA lies approximately between one and five. In [21], using insurance data from Israel, they report an average RRA of 97. In [47], using once again insurance data, the author finds an estimation of the RRA of the order of few thousands. As noted by the author, the lower bound on the observed RRA is around 1,000 times the level estimated by these existing studies. Following the aforementioned articles, we will conduct our simulations using a range of values for the RRA denoted $\rho$ from 1 to 50. This choice allows us to explore a broad spectrum of risk aversion levels while our upper bound is still lower than some found in the literature.

*Loss distribution.* In the literature on insurance, only few papers provide descriptive statistics on observed losses (claims size) or on the claim rate. For the numerical analysis, when the data are dimensionalized, we shall consider the coefficient of variation which is the dimensionless quantity defined as $v := \sqrt{\mathcal{V}}/E$. For instance, in [39], they consider a dataset that consists of 6773 amounts (in U.S. dollars) paid by a large US insurance company to settle and close claims for private passenger automobile policies. From their Table 7, the mean (claim size) is equal to 1853 and the standard deviation is equal to 2647 so that the coefficient of variation is equal to approximately 1.45. Since we consider the distribution of the type (probability of a claim) and not the distribution of the loss, it would be more appropriate to use descriptive statistics on the claim probabilities. Interestingly, in [4], see also [6], they report the predicted distribution of a claim probability. More specifically, they report the graph of the empirical density, the mean, the standard deviation of the predicted claim probability for three different insurances, auto collision, auto comprehensive and home. From their Table 4, the mean predicted probability is equal respectively to 0.069 for auto collision, 0.021 for auto comprehensive and 0.084 for home while the standard deviations are respectively equal to 0.024, 0.011 and 0.044. It thus follows that the coefficient of variation $v$ is approximately equal to 0.35 for auto collision and to (approximately) 0.525 for auto comprehensive and home. Regarding now the shape of the empirical density function for the predicted claim probabilities reported in [4], it is single-peaked and thus corresponds to a (Beta) density which is A-shaped (see Fig. C.9) within our framework.

The purpose of this discussion is to establish realistic exogenous parameters that will enable us to conduct numerical simulations within a range of values that are sensible and practical. In the subsequent sections, we will carefully select our exogenous parameters, namely $E$, $Z$, and $\rho$, in accordance with the observed stylized facts.

- Regarding risk aversion, we shall explore situations in which $\rho$ varies from say 1 to 50 to understand how optimal quantities evolve with $\rho$. A reasonable value for $\rho$ will be around 5.

- When selecting values for the parameters $E$ and $Z$, which determine $\sqrt{\mathcal{V}} := \sigma(E, Z)$, we will refer to the descriptive statistics provided in [4] as a useful benchmark. It's important to note that their approach involves households facing a menu of contracts with different premiums and deductibles, whereas in our framework, each agent faces a unique contract[9]. When considering a specific set of exogenous parameters $E$, $Z$, and $\rho$, we derive the optimal segmentation $\theta^*$, as well as the optimal indemnity $r^*$ and premium $\bar{p}(\theta^*, r^*, \rho)$. To make meaningful comparisons with the descriptive statistics in [4], we will not directly compare $E$ or $\sqrt{\mathcal{V}}$ with their data. Instead, we will compare $A(\theta^*)$ and $\sqrt{\mathcal{V}(\theta^*)}$, representing the mean and standard deviation conditional on $\theta \geq \theta^*$, since only the subset of agents $[\theta^*, 1]$ is insured.

*7.3. Should the Indemnity be Regulated?*

For this (prescriptive) analysis[10], we will select the three parameters, namely $E$, $Z$, and $\rho$, to align as closely as possible with the literature as mentioned in Section 7.2. Given a specific set of parameters, we can determine the optimal quantities $\theta^*$ and $r^*$ for the unregulated case in which $l = 1$. Let's consider the values $E = 0.05$ and $Z = 0.989$, leading to $\sqrt{\mathcal{V}} \approx 0.023$ and $v \approx 0.44$. With these values of $E$ and $Z$, the underlying Beta distribution is single-peaked. When $\rho = 5$ in the unregulated scenario, we find that $\theta^* \approx 0.0285$, $A(\theta^*) \approx 0.053$, and $\sqrt{\mathcal{V}}(\theta^*) = 0.02$. These results align with the observations made in [4].

---

[9]In [4], the assumption is that the choice of deductible does not influence the claim rate process (i.e., no moral hazard), which follows a Poisson process with intensity dependent on the household and coverage. This assumption aligns with our model, where the risk of damage for each agent is exogenous.

[10]The details of the implementation and some computation details are provided in Appendix Appendix D.

Figure 3 features a series of graphs to illustrate different scenarios. The $x$-axis of each graph represents the relative risk aversion (RRA) $\rho$, allowing us to observe the changes as $\rho$ varies in $(0, 50)$.

In the first column, we examine the scenario where indemnity is not regulated. Here, the insurer can freely determine $r$ and $\theta$ to maximize its profit. Starting from the top, the first graph displays the insurer's profit and the social welfare. Moving to the second graph, we observe the take-up rate, the optimal market segmentation $\theta^*$, the value $A(\theta^*)$, and the indemnity $r$. The third graph in this column illustrates the premium and the marginal cost. The bottom graph (fourth one) represents the elasticity.

The center column showcases the scenario with indemnity regulation. In this case, the regulator selects $r$ while anticipating the resulting $\theta_r^*$ chosen by the insurer, who aims to maximize its profit. The regulator's objective is to choose $r$ to maximize social welfare. The second column presents graphs for all the variables mentioned earlier in this regulated situation.

To facilitate comparison between the unregulated situation (left column) and the regulated scenario (center column), we provide a relative increase (or decrease) from the left to the center column. This comparative information is presented in the right column. When the variable being examined is very close to zero, the comparison graph in the right column may exhibit noise. It is the case for the marginal cost. Nevertheless, the overall outcome remains unaffected.

The observation of Figure 3 highlights that both in the regulated and unregulated cases, the shapes of the graphs are similar. Several notable patterns emerge from the graphs (Figure 3) that hold true regardless of whether the insurer is regulated or not:

- The optimal segmentation, average probability, and marginal cost display relatively constant behavior as $\rho$ varies.

- The optimal premium, optimal indemnity, and resulting take-up rate exhibit increasing (and concave) trends as $\rho$ increases. However, for values of $\rho$ exceeding ten, both the optimal indemnity and take-up rate become relatively stable.

- Both the profit and the social welfare demonstrate increasing (and concave) trends as $\rho$ increases.

From these distinct patterns, it becomes evident that the monopolist markup, defined as the difference between the optimal premium and the average cost, is positively correlated with $\rho$. This implies that risk aversion can serve as a measure of market power: greater risk aversion corresponds to a higher level of market power for the monopolist. Given that the take-up rate also follows an increasing (but concave) trend with $\rho$, the total profit of the monopolist increases with $\rho$ as both the average profit and the quantity of contracts rise.

An interesting observation is that, for a fixed $E$, both the parameters $\rho$ and $Z$ have a similar impact on the profit, but they exert opposite effects on the optimal premium. Specifically, the optimal premium increases with $\rho$ while decreasing with $Z$.

To compare the unregulated case (variables subscripted with "Reg") with the regulated one (variables subscripted with "Unreg"), we present the percentage differences of the optimal quantities with respect to the unregulated case, using the latter as the benchmark. Assuming $\rho = 5$, Table 1 displays the percentage variations in relation to $\rho$. For instance, in the table,

$$\frac{r^*_{\text{Reg}} - r^*_{\text{Unreg}}}{r^*_{\text{Unreg}}} = +11.2\% \qquad \text{and} \qquad \frac{Q^*_{\text{Reg}} - Q^*_{\text{Unreg}}}{Q^*_{\text{Unreg}}} = -0.4\%.$$

The numerical results make it evident that by regulating the monopolistic insurer (i.e., selecting the indemnity to optimize social welfare), the indemnity can be significantly increased, while other quantities remain relatively unaffected. Here are the key findings for different levels of risk aversion:

- When $\rho = 1$, the indemnity increases by 15%, but the premium also increases by 13%. This result suggests that the impact of regulation is not entirely clear in terms of its benefit.

- For $\rho = 5$, the indemnity increases by more than 11%, whereas the premium only increases by 1.7%, and the take-up rate remains essentially unchanged.

- When $\rho = 10$, the indemnity increases by around 10%, with a premium increase of less than 1%, and the take-up rate showing minimal variation.

In summary, our numerical simulations indicate that the proposed monopoly regulation outlined in this study becomes especially relevant when risk aversion is relatively high, exceeding 5. This is due to the fact that such regulation can lead to a substantial increase in the quality of the contract (indemnity), while maintaining stability in other quantities.

| $\rho$ | Indemnity (in %) | Take up rate (in %) | Profit (in %) | Premium (in %) |
|---|---|---|---|---|
| 0.5 | +16.0 % | -8.3 % | -1.8 % | +16.9 % |
| 1.0 | +15.1 % | -5.0 % | -1.6 % | +13.3 % |
| 2.5 | +12.1 % | -1.4 % | -0.7 % | +5.1 % |
| 5.0 | +11.2 % | -0.4 % | -0.4 % | +1.7 % |
| 7.5 | +10.3 % | -0.2 % | -0.3 % | +0.9 % |
| 10.0 | +9.4 % | -0.1 % | -0.2 % | +0.6 % |

Table 1: Comparison of the unregulated vs. the regulated case. $E = 5\%$ (mean) and $v = 0.445$ (coefficient of variation)

A natural question arises: would this observation hold for different values of $E$ and $Z$? To answer this question, Figure 4 presents a comprehensive numerical exploration of the entire $EZ$-square, investigating all possible Beta distributions. For every combination of $(E, Z)$, we indicate the relative increase of the indemnity in scenarios where indemnity is unregulated compared to instances where it is. The graphical representation employs colors of the visible spectrum, where purple corresponds to no increase and red signifies a 54% increase.

This thorough analysis not only corroborates the observation outlined above but also establishes its general applicability, extending beyond specific choices of $E$ and $Z$. Additionally, it pinpoints where (i.e. for which type of agent risks) the increase is more important.

## 8. Conclusion

In this study, we have investigated the theoretical foundations of community rating by a private insurer in the absence of an individual mandate, under the constraint of designing a single insurance contract. Our investigation has revealed that, under certain conditions, the optimal premium aligns with the Lerner index. We have also decomposed the elasticity of demand into two components, one associated with the market and the other linked to risk aversion.

Regarding community rating regulation, our work provides insight on key variables such as the optimal contract design (indemnity and premium) and the take-up rate. Our numerical findings suggest that in the absence of indemnity regulation, its quality will be lower but it highlights the importance of regulating the insurer monopolist when risk aversion is high.

We anchored our community rating framework within Stiglitz's model (1977), which considers a continuum of agents differentiated solely by their probability of incurring damage. Future extensions could explore scenarios where agents also differ in terms of risk aversion, enabling an analysis of advantageous selection situations where damage probability correlates with risk aversion. Another avenue could involve examining cases where the insurer is constrained by regulation to offer a specific number of contracts. This might lead to higher take-up rates and more profound extraction of policyholders' surplus due to increased pricing flexibility. However, the net effect on social welfare in comparison to the one-contract benchmark remains uncertain. The finite contracts scenario presents intriguing theoretical challenges as it introduces participation and incentive constraints for each contract and market segment.

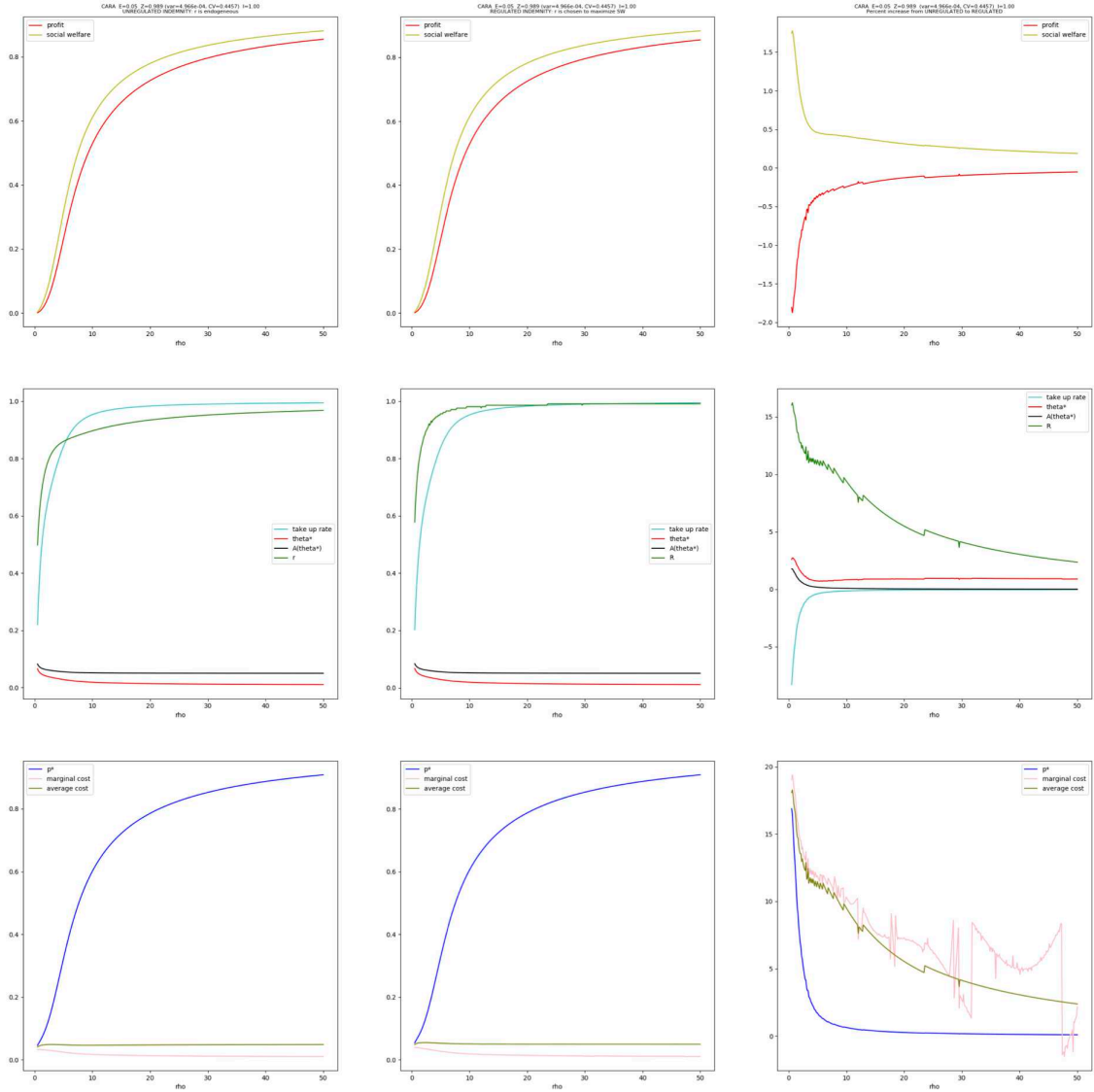

Figure 3: Profit, social welfare, take-up rate, optimal $\theta^*$, $A(\theta^*)$, $r$, premium, marginal cost, and elasticity, all for $E = 0.05$, $Z = 0.989$, $v = 0.445$. The left column represents the unregulated indemnity scenario, while the center column showcases the regulated indemnity case. The right column illustrates relative changes, facilitating comparison between the unregulated and regulated conditions.

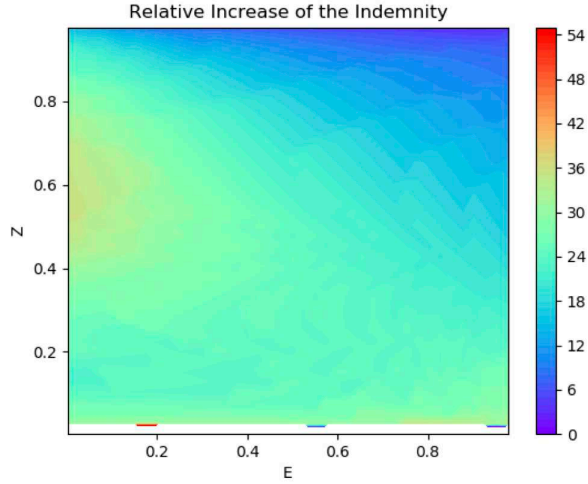

Figure 4: Comprehensive exploration of the the entire $EZ$-square for the relative increase of the indemnity in scenarios where indemnity is unregulated compared to instances where it is. RRA is $\rho = 5$.

**Appendix A. Proofs**

The aim of this appendix is to provided proofs to some of the results presented in this paper.

**Proof of Lemma 1** — Assume that there exists $\theta \in [0, 1)$ for which the contract $C$ satisfies the participation constraint (3). This leads to the inequality

$$\theta[U(W_0 - L - P + R) - U(W_0 - L)] + (1 - \theta)[(U(W_0 - P) - U(W_0)] \geq 0.$$

Given that $U$ is an increasing function and $P > 0$, it is clear that $U(W_0 - P) - U(W_0) < 0$. Since $\theta < 1$, we can deduce that $U(W_0 - L - P + R) - U(W_0 - L)$ must be positive. Using the fact that $U$ is an increasing function, we can then deduce that $P < R$.

Conversely, assume that $C = (P, R)$ is such that $P < R$. Solving $V(\theta, C) = V(\theta, 0)$ for $\theta$ gives one solution denoted $\bar{\theta}(P, R)$, given by

$$\bar{\theta}(P, R) = \frac{U(W_0) - U(W_0 - P)}{[U(W_0) - U(W_0 - P)] + [U(W_0 - L + R - P) - U(W_0 - L)]} \tag{A.1}$$

We shall call it the critical value. Using $R - P > 0$ and $U$ increasing gives $U(W_0 - L + R - P) - U(W_0 - L) > 0$, which gives $\bar{\theta}(P, R) < 1$. Therefore agents of the group $[\bar{\theta}(P, R), 1]$ purchase the contract, which indicate that $C$ meets the participation constraint. We call $\bar{\theta}(P, R)$ the critical value since the behavior if the agents changes at this point. □

**Proof of Proposition 1** — Three parts have to be proven.

($i$) Let $R \in [0, L]$, then

- $\mathcal{G}(0, P, R) = 0$ is equivalent to $U(W_0 - P) = U(W_0)$, which is equivalent to $P = 0$, thanks to the strict monotonicity of $U$. Thus $\forall R \in [0, L]$, $\bar{P}(0, R) = 0$.

- Similarly, $\mathcal{G}(1, P, R) = 0$ is equivalent to $U(W_0 - L - P - R) = U(W_0 - R)$, which is in turn equivalent to $P = R$.

Not surprisingly, an agent with no risk will chose not to buy insurance ($P(0, R) = 0$) and an agent sure to have a damage, will be willing to pay a premium up to the indemnity ($P(1, R) = R$).

(ii) From the definition of function $\mathcal{G}$ in (5), and the regularity that derives from the regularity of $U$, we have:

$$\frac{\partial \mathcal{G}}{\partial \theta}(\theta, P, R) = U(W_0 - L - P + R) - U(W_0 - P) - U(W_0 - L) + U(W_0) > 0 \tag{A.2}$$

$$\frac{\partial \mathcal{G}}{\partial P}(\theta, P, R) = -\theta U'(W_0 - L - P - R) - (1 - \theta)U'(W_0 - P) < 0 \tag{A.3}$$

$$\frac{\partial \mathcal{G}}{\partial R}(\theta, P, R) = \theta U'(W_0 - L - P - R) > 0 \tag{A.4}$$

The Implicit Function Theorem yields:

$$\frac{\partial \bar{P}}{\partial \theta}(\theta, R) = -\frac{\frac{\partial \mathcal{G}}{\partial \theta}(\theta, \bar{P}(\theta, R), R)}{\frac{\partial \mathcal{G}}{\partial P}(\theta, \bar{P}(\theta, R), R)} \tag{A.5}$$

Substituting (A.2) and (A.4) in (A.5) gives $\frac{\partial \bar{P}}{\partial \theta}(\theta, R) > 0$. Similarly, an analogous reasoning can be employed to demonstrate that $\frac{\partial \bar{P}}{\partial R}(\theta, R) > 0$.

(iii) From (A.5) and the regularity derived from $U$, we can compute:

$$\frac{\partial^2 \bar{P}}{\partial \theta^2}(\theta, R) = -\frac{\frac{\partial}{\partial \theta}\left(\frac{\partial \mathcal{G}}{\partial \theta}(\theta, \bar{P}(\theta, R), R)\right)\frac{\partial \mathcal{G}}{\partial P}(\theta, \bar{P}(\theta, R), R) - \frac{\partial \mathcal{G}}{\partial \theta}(\theta, \bar{P}(\theta, R), R)\frac{\partial}{\partial \theta}\left(\frac{\partial \mathcal{G}}{\partial P}(\theta, \bar{P}(\theta, R), R)\right)}{\left(\frac{\partial \mathcal{G}}{\partial P}(\theta, \bar{P}(\theta, R), R)\right)^2} \tag{A.6}$$

Given:

- The independence of $\frac{\partial \mathcal{G}}{\partial \theta}$ with respect $\theta$, which leads to $\frac{\partial^2 \mathcal{G}}{\partial \theta^2}(\theta, \bar{P}(\theta, R), R) = 0$.
- The result established in (ii) providing $\frac{\partial \bar{P}}{\partial \theta}(\theta, R) \geq 0$.
- The decrease of $U'$, which leads to $\frac{\partial^2 \mathcal{G}}{\partial \theta \partial P}(\theta, \bar{P}(\theta, R), R) = U'(W_0 - P) - U'(W_0 - P - (L - R)) \leq 0$.

We have:

$$\frac{\partial}{\partial \theta}\left(\frac{\partial \mathcal{G}}{\partial \theta}(\theta, \bar{P}(\theta, R), R)\right) = \frac{\partial^2 \mathcal{G}}{\partial \theta^2}(\theta, \bar{P}(\theta, R), R) + \frac{\partial \bar{P}}{\partial \theta}(\theta, R)\frac{\partial^2 \mathcal{G}}{\partial \theta \partial P}(\theta, \bar{P}(\theta, R), R) \leq 0$$

Using (A.3), we have:

$$\frac{\partial}{\partial \theta}\left(\frac{\partial \mathcal{G}}{\partial \theta}(\theta, \bar{P}(\theta, R), R)\right)\frac{\partial \mathcal{G}}{\partial P}(\theta, \bar{P}(\theta, R), R) \geq 0 \tag{A.7}$$

Moreover:

- If $U''$ is increasing, then $U''(W_0 - \bar{P}(\theta, R) - (L - R)) - U''(W_0 - \bar{P}(\theta, R)) \leq 0$. If $R = L$ the expression vanishes. In any case, it is non-positive.
- The concavity of $U$ yields that $U''(W_0 - \bar{P}(\theta, R))$ is non-positive.

therefore

$$\frac{\partial^2 \mathcal{G}}{\partial P^2}(\theta, \bar{P}(\theta, R), R) = \theta[U''(W_0 - \bar{P}(\theta, R) - (L - R)) - U''(W_0 - \bar{P}(\theta, R))] + U''(W_0 - \bar{P}(\theta, R)) \leq 0$$

Since $\frac{\partial^2 \mathcal{G}}{\partial \theta \partial P}(\theta, \bar{P}(\theta, R), R) \leq 0$ and $\frac{\partial \bar{P}}{\partial \theta}(\theta, R) \geq 0$ as explained earlier, we have:

$$\frac{\partial}{\partial \theta}\left(\frac{\partial \mathcal{G}}{\partial P}(\theta, \bar{P}(\theta, R), R)\right) = \frac{\partial^2 \mathcal{G}}{\partial P \partial \theta}(\theta, \bar{P}(\theta, R), R) + \frac{\partial \bar{P}}{\partial \theta}(\theta, R)\frac{\partial \mathcal{G}^2}{\partial P^2}(\theta, \bar{P}(\theta, R), R) \leq 0$$

Using $\frac{\partial \mathcal{G}}{\partial \theta}(\theta, P, R) > 0$. it follows:

$$\frac{\partial \mathcal{G}}{\partial \theta}(\theta, \bar{P}(\theta, R), R) \frac{\partial}{\partial \theta}\left(\frac{\partial \mathcal{G}}{\partial P}(\theta, \bar{P}(\theta, R), R)\right) \leq 0 \tag{A.8}$$

From (A.7) and (A.8), the numerator of the right hand side of (A.6) is non-negative. Its denominator is obviously positive. Subsequently,

$$\frac{\partial^2 \bar{P}}{\partial \theta^2}(\theta, R) \leq 0$$

which proves the concavity. Using $(i)$, we obtain $\bar{P}(\theta, R) \geq \theta R$.

$\square$

**Proof of lemma 3** — Let us prove $(i)$ and $(iv)$.

$(i)$ Let $\theta \in (0, 1)$, then for all $x \in [\theta, 1)$ we have $f(x) > \theta f(x)$. By integrating over $[\theta, 1]$ and dividing by $1 - F(\theta)$ we obtain

$$A(\theta) > \frac{1}{1 - F(\theta)} \int_{\theta}^{1} \theta f(x)\, dx.$$

The right hand side simplies to $\theta$ thus $A(\theta) > \theta$. The proof of $m(\theta) > 0$ for $\theta \in (0, 1)$ derives directly from $A(\theta) > \theta$.

$(iv)$ Let $\theta \in (0, 1)$. From (16) we have $A''(\theta) = H'(\theta)m(\theta) + H(\theta)m'(\theta)$. Since $m(\theta) > 0$, we have:

$$H'(\theta) = \frac{1}{m(\theta)}\left[A''(\theta) + H(\theta)(1 - A'(\theta))\right] \tag{A.9}$$

Given that $A'$ is increasing on $(0, 1]$ and $A'(1) < 1$, it follows that $A'(\theta) < 1$ for all $\theta \in (0, 1]$. Consequently, for all $\theta \in (0, 1)$, we have $1 - A'(\theta) > 0$. We also have $m(\theta) > 0$ and $H(\theta) > 0$. By hypothesis, we have $A''(\theta) > 0$. By examining the sign of each term in equation (A.9), we deduce the result.

Of course $A''(\theta) \geq 0$ serves as a sufficient condition, it is not necessary. This is because a negative value of $A''(\theta)$ can potentially be compensated in (A.9) by a positive value of $H(\theta)(1 - A'(\theta))$, where the latter term is greater than that the absolute value of $A''(\theta)$.

$\square$

**Proof of lemma 4** — Consider now the margial cost, denoted MC$(\theta, R)$. By definition, it provides the variation of total cost TC with respect to the change in quantity resulting from a shift in $\theta$. The function $\theta \mapsto Q(\theta)$ is a continuous and (strictly) monotonic from $[0, 1]$ to $[0, 1]$, one thus can define the inverse mapping $Q^{-1}$. Let us define the auxiliary function $\overline{\text{TC}}$ on $[0, 1] \times [0, L]$:

$$\overline{\text{TC}}(q, R) = R \cdot A(Q^{-1}(q)) \cdot q \tag{A.10}$$

Where $q$ represents a quantity, so that $\overline{\text{TC}}(Q(\theta), R) = \text{TC}(\theta, R)$. The derivative of $\overline{\text{TC}}$ with respect to $q$ gives the marginal cost. Through simple calculus, we obtain:

$$\frac{\partial \overline{\text{TC}}}{\partial q}(q, R) = (Q^{-1})'(q) A'(Q^{-1}(q))\, R\, q + A(Q^{-1}(q))\, R = \frac{A'(Q^{-1}(q))}{Q'(Q^{-1}(q))} R\, q + A(Q^{-1}(q))\, R.$$

Subsequently

$$\text{MC}(\theta, R) = \frac{\partial \overline{\text{TC}}}{\partial q}(Q^{-1}(\theta), R) = \frac{A'(\theta) Q(\theta)}{Q'(\theta)} \cdot R + A(\theta) \cdot R \tag{A.11}$$

This equation can be simplified by applying (15) and 16. Substituting the appropriate expressions leads to

$$\mathrm{MC}(\theta, R) = \theta \cdot R \tag{A.12}$$

Using Lemma 3 (i) and the expressions in (19) and (A.12), we can establish an order between the average cost and the marginal cost.□

**Proof of Fact 5** — Without loss of generality, risk-neutrality corresponds to a utility function is $U(W) = W$ for all $W \in [0, W_0]$. From (5), we have $\mathcal{G}(\theta, P, R) = \theta R - P$. It follows that $\bar{P}(\theta, R) = \theta R$. Subsequently, (9) gives for all $(\theta, R) \in [0, 1] \times [0, L]$:

$$\Pi(\theta, R) = \int_\theta^1 (\theta R - xR)\, d\mu(x) = \theta R\,[1 - F(\theta)] - R\,A(\theta)\,[1 - F(\theta)] = R\,[1 - F(\theta)]\,(\theta - A(\theta)) \leq 0$$

proving no contract is profitable. □

**Proof of Proposition 2** — Let us consider the set

$$S = \{x \in [0, 1],\ \mathrm{s.t}, \mathcal{R}([x, 1)) \subset (0, +\infty)\}$$

Since $A$ is differentiable in 1, we have

$$\frac{\partial \mathcal{R}}{\partial \theta}(\theta, R) = \frac{\partial \bar{P}}{\partial \theta}(\theta, R) - R\,A'(\theta).$$

In $\theta = 1$, the assumption (24) yields $\frac{\partial \mathcal{R}}{\partial \theta}(1, R) < 0$. Since $\mathcal{R}(1, R) = 0$, thre exists $\varepsilon > 0$ such that $\mathcal{R}$ positive on $[1 - \varepsilon, 1)$. Subsequently, $1 - \varepsilon \in S$. Now, $S$ is a non-empty bounded set and we define $\hat{\theta}_R = \inf S$. □

**Proof of Lemma 5** — Assume that $\hat{\Theta}(R)$ is not empty so that there exists at least one optimal profitable segmentation $\theta^* \in \hat{\Theta}(R)$. For simplicity, we shall write the optimal segmentation as $\theta^*$ instead of $\theta_R^*$.

First, let us prove equation (26a). We have $Q(\theta) = 1 - F(\theta)$, thus $Q'(\theta) = -f(\theta) < 0$. Using (22), we can differentiate $\Pi(\theta, R)$ with respect to $\theta$. We obtain

$$\frac{\partial \Pi}{\partial \theta}(\theta, R) = Q(\theta)\left(\frac{\partial \bar{P}}{\partial \theta}(\theta, R) - R\,A'(\theta)\right) - f(\theta)(\bar{P}(\theta, R) - R\,A(\theta)) \tag{A.13}$$

For a set $R$, the necessary conditions of optimality $\frac{\partial \Pi}{\partial \theta}(\theta^*, R) = 0$, lead to

$$Q(\theta^*)\left[\frac{\partial \bar{P}}{\partial \theta}(\theta^*, R) - R\,A'(\theta^*)\right] = f(\theta^*)[\bar{P}(\theta^*, R) - R\,A(\theta^*)] \tag{A.14}$$

Using that $H(\theta) = f(\theta)/(1 - F(\theta))$ and that both $f(\theta)$ and $1 - F(\theta)$ are positive for all $\theta \in [0, 1)$, we have

$$\frac{\partial \bar{P}}{\partial \theta}(\theta^*, R) - R\,A'(\theta^*) = H(\theta^*)[\bar{P}(\theta^*, R) - R\,A(\theta^*)] \tag{A.15}$$

By definition of an optimal profitable segmentation, $\theta^* \in \hat{\Theta}(R)$, $\Pi(\theta^*, R) > 0$ so that $\mathcal{R}(\theta^*, R) = \bar{P}(\theta^*, R) - R\,A(\theta^*) > 0$. Since $H(\theta^*) > 0$, it thus follows that $\frac{\partial \bar{P}}{\partial \theta}(\theta^*, R) - R\,A'(\theta^*) > 0$ so that the optimal segmentation $\theta^*$ satisfies (26a).

Let us now prove (26b). Let us introduce $G$, an anti-derivative of $x \mapsto R\,x\,f(x)$. From (22), we have

$$\Pi(\theta, R) = \underbrace{\bar{P}(\theta, R)\,[1 - F(\theta)]}_{\text{Revenue}} - \underbrace{[G(1) - G(\theta)]}_{\text{Cost}} \tag{A.16}$$

therefore
$$\frac{\partial \Pi}{\partial \theta}(\theta, R) = Q(\theta)\frac{\partial \bar{P}}{\partial \theta}(\theta, R) - f(\theta)(\bar{P}(\theta, R)) + R\theta f(\theta)$$

Again, for a given $R$, the necessary optimality condition $\frac{\partial \Pi}{\partial \theta}(\theta^*, R) = 0$ leads to

$$Q(\theta^*)\frac{\partial \bar{P}}{\partial \theta}(\theta^*, R) = f(\theta^*) \times [(\bar{P}(\theta^*, R)) - R\theta^*] \tag{A.17}$$

We already know that $[(\bar{P}(\theta^*, R)) - R\theta^*] > 0$. Using the definition of the hazard rate evaluated in $\theta^*$, $H(\theta^*) = \frac{f(\theta^*)}{1-F(\theta^*)} > 0$, equation (A.17) leads to (26b). □

**Proof of proposition 3** — The proof will be structured as follows: we will outline three claims, and subsequently use these claims to establish the theorem.

*Claim 1:* If $U''$ is increasing, $\mu \in \mathcal{M}_3$ and $R\,A'(1) > \frac{\partial \bar{P}}{\partial \theta}(1, R)$, then $\hat{\Theta}(R) = (\hat{\theta}_R, 1)$.

Let us prove this first claim. Given that $R\,A'(1) > \frac{\partial \bar{P}}{\partial \theta}(1, R)$, Lemma 2 gives $\hat{\Theta}(R) \neq \emptyset$. With $U''$ being increasing, using Proposition 1 (iii), it follows that $\theta \mapsto \bar{P}(\theta, R)$ is concave. Furthermore, due to $\mu \in \mathcal{M}_3$, the function $A''$ remains non-negative within the interval $(0, 1)$, consequently $-R \times A$ is concave. As a result, $\theta \mapsto \mathcal{R}(\theta, R) = \bar{P}(\theta, R) - RA(\theta)$ is concave. Since $\mathcal{R}(0, R) < 0$ and $\mathcal{R}(1, R) = 0$, there is unique $\hat{\theta}_R \in (0, 1)$ such that for any $\theta \in (\hat{\theta}_R, 1)$, $\mathcal{R}(\theta, R) > 0$, implying $\hat{\Theta}(R) = (\hat{\theta}_R, 1)$, which concludes de proof of this first claim.

*Claim 2:* If $\mu \in \mathcal{M}_3$, then $\mathcal{R}$ is log-concave on $\hat{\Theta}(R) = (\hat{\theta}_R, 1)$.

Let us prove this second claim. First note that

$$\frac{\partial^2}{\partial \theta^2}[\ln(\mathcal{R}(\theta, R)] = \frac{\frac{\partial^2 \mathcal{R}}{\partial \theta^2}(\theta, R)\mathcal{R}(\theta, R) - (\frac{\partial \mathcal{R}}{\partial \theta}(\theta, R))^2}{(\mathcal{R}(\theta, R))^2} \tag{A.18}$$

Therefore the concavity of $\mathcal{R}$ will be driven by the numerator appearing on the right hand side of (A.18). Let us compute the numerator:

$$\frac{\partial^2 \mathcal{R}}{\partial \theta^2}(\theta, R)\mathcal{R}(\theta, R) = \left(\frac{\partial^2 \bar{P}}{\partial \theta^2}(\theta, R) - R\,A''(\theta)\right)(\bar{P}(\theta, R) - RA(\theta)) \tag{A.19}$$

Since $U''$ increasing and $\mu \in \mathcal{M}_3$, we thus have $\frac{\partial^2 \bar{P}}{\partial \theta^2}(\theta, R) - R\,A''(\theta) \leq 0$. Moreover, $(\bar{P}(\theta, R) - R\,A(\theta)) > 0$ for $\theta \in (\hat{\theta}_R, 1)$, it thus follows the right hand side of (A.19) is negative so that the right hand side of (A.18) is also negative, and this concludes the proof of the second claim.

*Claim 3:* If $\mu \in \mathcal{M}_3$, then $\theta_R^* \in (\hat{\theta}_R, 1)$ is unique.

To prove this third claim, let us consider the optimality condition given by (26a) that we now call $g_R$. It is defined on $(0, 1)$ and given by

$$g_R(\theta) = \frac{\frac{\partial \bar{P}}{\partial \theta}(\theta, R) - R\,A'(\theta)}{\bar{P}(\theta, R) - R\,A(\theta))} - H(\theta) \tag{A.20}$$

We have $g_R(\theta) = \frac{\partial}{\partial \theta}[\ln(\mathcal{R}(\theta, R)] - H(\theta)$ therefore

$$g_R'(\theta) = \frac{\partial^2}{\partial \theta^2}[\ln(\mathcal{R}(\theta, R)] - H'(\theta)$$

In Claim 2, we have established that $\frac{\partial^2}{\partial \theta^2}[\ln(\mathcal{R}(\theta, R)] < 0$ and Lemma 3 (iv) gives $H'(\theta) > 0$ on $(0, 1)$. It follows $g_R$ is strictly monotonic decreasing. As a result, (26a) possesses, at most, one solution. Since we have already established the existence of one solution, the proof for claim 3 is thereby concluded.

We are now in a position to prove that the optimal premium satisfies the so-called inverse elasticity rule. Let us rewrite (26b) as

$$\bar{P}(\theta_R^*, R) - R\,\theta_R^* = \frac{\frac{\partial \bar{P}}{\partial \theta}(\theta_R^*, R)}{H(\theta_R^*)}.$$

Using the fact that marginal cost $\mathrm{MC}(\theta_R^*, R)$ is equal to $R\,\theta_R^*$ and that $H = -Q'/Q$, the previous equation writes:

$$\bar{P}(\theta_R^*, R) - \mathrm{MC}(\theta_R^*, R) = -\frac{\frac{\partial \bar{P}}{\partial \theta}(\theta_R^*, R) Q(\theta_R^*)}{Q'(\theta_R^*)}.$$

From the definition of the elasticity of demand in equation (27), it follows that

$$\bar{P}(\theta_R^*, R) - \mathrm{MC}(\theta_R^*, R) = -\frac{\bar{P}(\theta_R^*, R)}{\epsilon(\theta_R^*)}.$$

so that

$$\frac{\bar{P}(\theta_R^*, R) - \mathrm{MC}(\theta_R^*, R)}{\bar{P}(\theta_R^*, R)} = -\frac{1}{\epsilon(\theta_R^*)}$$

which can be rearranged to

$$\bar{P}(\theta_R^*, R) \times \left(1 + \frac{1}{\epsilon(\theta_R^*)}\right) = \mathrm{MC}(\theta_R^*, R)$$

which is (29). Furthermore, since both the premium $\bar{P}(\theta_R^*, R)$ and $\bar{P}(\theta_R^*, R) - \mathrm{MC}(\theta_R^*, R)$ are positive, we have

$$0 < 1 + \frac{1}{\epsilon(\theta_R^*)} < 1 \tag{A.21}$$

Using $\epsilon(\theta_R^*) < 0$, since $1 + \frac{1}{\epsilon(\theta_R^*)} > 0$ we derive from (A.21) that $\epsilon(\theta_R^*) < -1$, i.e., $|\epsilon(\theta_R^*)| > 1$. $\square$

## Appendix B. The Regularity of the Average Probability of Damage function

The Average Probability of Damage function $A$ plays a significant role in the profit resulting from a market and subsequently the profitability of the said market. The aim of the appendix is to study the regularity and main properties of this function.

*Appendix B.1. Continuity and Definition in 1*

**Lemma 7.** *The function $A$ defined on $[0, 1)$ by equation (13) can be extended by continuity in 1 to $A(1) = 1$.*

*Proof —* Let $x \in (0, 1)$. Consider $\theta \in [x, 1[$ then $xf(\theta) \leq \theta f(\theta) \leq f(\theta)$ thus

$$x \int_x^1 f(\theta) d\theta \leq \int_x^1 \theta f(\theta) \, d\theta \leq \int_x^1 f(\theta) \, d\theta$$

Dividing now by $\int_x^1 f(\theta) d\theta = 1 - F(x)$ yields $x \leq A(x) \leq 1$ therefore

$$\lim_{x \to 1} A(x) = 1.$$

Since the limit of $A$ is finite in 1, we can extend the function by continuity. $\square$

*Appendix B.2. Differentiability*

Having extended the function $A$ by continuity in 1, it is now defined and continuous on the entire interval $[0, 1]$. However, the differentiability for $A$ can only be guaranteed on the interval $[0, 1)$. For $A$ to be differentiable in one, need to have conditions on the density $f$, however these turn out to be very weak.

Before we define our "docking index", we shall recall the equivalence of two functions $f_1$ and $f_2$ at a given point $x_0$, denoted as $f_1 \sim_{x_0} f_2$, means there exists of a function $g$ such that $f_1(x) = (1 + g(x))f_2(x)$, with $\lim_{x \to x_0} g(x) = 0$. In most scenarios, i.e. non-pathological cases, this simply means the ratio $f_1(x)/f_2(x)$ converges towards 1 as $x$ approaches $x_0$.

**Definition 6.** *A density function $f$ admits a "docking index", which will be denoted $\mathrm{d}(f)$, if it is equivalent to a power function in 1. In this case we define the docking index as the exponent of the power function.*

The docking index quantifies how closely the density function aligns or "docks" at the critical point 1, where the agent will almost surely suffer damage leading to a claim. A positive and high docking index indicates that the probability distribution exhibits relative flatness as it approaches 1, so that highly risky agents are spread out around 1.

**Lemma 8.** *Let $f$ be a function admitting a docking index $\mathrm{d}(f) > -1$, then $A$ is differentiable at 1 and*

$$A'(1) = \frac{\mathrm{d}(f) + 1}{\mathrm{d}(f) + 2} \tag{B.1}$$

*Proof —* Let $h \in (0, 1)$ and consider $x = 1 - h$. From (13), the average probability evaluated at $1 - h$ is

$$A(1 - h) = \frac{\int_{1-h}^{1} \theta f(\theta) d\theta}{1 - F(1 - h)} \tag{B.2}$$

Using $A(1) = 1$, we have

$$A(1 - h) - A(1) = \frac{\int_{1-h}^{1} (\theta - 1) f(\theta) \, d\theta}{\int_{1-h}^{1} f(\theta) \, d\theta}$$

Since $f$ admits a docking index, there exist $s > -1$ and $c > 0$ such that $f(\theta) \sim_1 c(1 - \theta)^s$. That is

$$f(\theta) = c(1 - \theta)^s (1 + g(\theta))$$

with $\lim_{\theta \to 1} g(\theta) = 0$. Therefore

$$A(1 - h) - A(1) = -\frac{\int_{1-h}^{1} (1 + g(\theta)) (1 - \theta)^{s+1} d\theta}{\int_{1-h}^{1} (1 + g(\theta)) (1 - \theta)^s d\theta} \tag{B.3}$$

Let $\epsilon \in (0, 2\frac{s+1}{s+2})$. Define $\varepsilon = \frac{\epsilon(s+2)}{2(s+1) - \epsilon(s+2)}$ so that we have $\epsilon = \frac{2\varepsilon}{1-\varepsilon} \frac{s+1}{s+2}$. Since $\lim_{x \to 1} g(x) = 0$, there exists $\eta > 0$ such that $h < \eta$ implies $\forall \theta \in (h, 1)$, $|g(\theta)| < \varepsilon$, which implies

$$1 - \varepsilon < 1 + g(\theta) < 1 + \varepsilon$$

In turn, it implies

$$\frac{1}{h} \frac{1 - \varepsilon}{1 + \varepsilon} \frac{\int_{1-h}^{1} (1 - \theta)^{s+1} d\theta}{\int_{1-h}^{1} (1 - \theta)^s d\theta} < \frac{A(1 - h) - A(1)}{-h} < \frac{1}{h} \frac{1 + \varepsilon}{1 - \varepsilon} \frac{\int_{1-h}^{1} (1 - \theta)^{s+1} d\theta}{\int_{1-h}^{1} (1 + g(\theta)) (1 - \theta)^s d\theta}$$

the integrals compute explicitly, giving:

$$\frac{1}{h} \frac{1 - \varepsilon}{1 + \varepsilon} \frac{h^{s+2}/(s+2)}{h^{s+1}/(s+1)} < \frac{A(1 - h) - A(1)}{-h} < \frac{1}{h} \frac{1 + \varepsilon}{1 - \varepsilon} \frac{h^{s+2}/(s+2)}{h^{s+1}/(s+1)}$$

therefore
$$\frac{1-\varepsilon}{1+\varepsilon}\frac{s+1}{s+2} < \frac{A(1-h)-A(1)}{-h} < \frac{1+\varepsilon}{1-\varepsilon}\frac{s+1}{s+2}$$

which is
$$\left|\frac{A(1-h)-A(1)}{-h} - \frac{s+1}{s+2}\right| < \epsilon$$

Hence
$$\forall \epsilon > 0,\ \exists \eta > 0,\ |h| < \eta \Rightarrow \left|\frac{A(1-h)-A(1)}{-h} - \frac{s+1}{s+2}\right| < \epsilon$$

which is the desired result. □

**Remark 3.** *If $f \in C^n([0,1])$, i.e., $f$ is n-times continuously differentiable on the interval $[0,1]$, we denote $f^{(0)}(1)$ the value of $f$ at $x = 1$, and for any integer $p \in [1,n]$, we denote $f^{(p)}(1)$ the $p^{th}$ derivative of the density function $f$ at $x = 1$. Let $\mathbb{D}$ be the set of integers $p < n$ for which $f^{(p)}(1) \neq 0$. If the set $\mathbb{D}$ is not empty, then a simple Taylor expansion of $f$ proves it admits a docking index and that $\mathrm{d}(f) = \min \mathbb{D}$, which in this case is necessarily an integer.*

*Appendix B.3. The Case of a Two-Parameter Beta Distribution*

For a Two-Parameter Beta distribution as the underlying probability measure, the differentiability of $A$ at 1 is assured, regardless of the chosen values for the parameters $\alpha$ and $\beta$ within the interval $(0, +\infty)$. This range of parameters also covers the scenario where $\beta \in (0,1)$, resulting in a discontinuity of $f$ at 1. This observation follows as a direct outcome of Lemma 8, with the observation that

$$f(x) \sim_1 \frac{1}{B(\alpha, \beta)}(1-\theta)^{\beta-1}$$

and consequently, $\mathrm{d}(f) = \beta - 1$. The essence of this is succinctly encapsulated in this corollary derived from Lemma 8. Note that a direct proof of the differentiability could also be done.

**Corollary 2.** *For a two-parameter Beta distribution of parameters $\alpha \in (0, +\infty)$ and $\beta \in (0, +\infty)$, the average probability of damage $A$ is differentiable in 1 and*

$$A'(1) = \frac{\beta}{\beta+1}$$

Our attention has been concentrated on the differentiability of $A$ at the upper endpoint of the interval $[0,1]$. This focus is motivated by the economic significance associated with this derivative. However, it is worthwhile to explore the differentiability of $A$ at the lower endpoint of the interval, namely, at 0. This is carried out in the subsequent two lemmas.

**Lemma 9.** *Consider a two parameter Beta distribution associated to a point in regions **A** or **I** of the EZ-square (that is $\mu = \mathrm{Beta}(\alpha, \beta)$ with $\alpha > 1$). Then $A$ is differentiable in 0 and*

$$A'(0) = 0 \tag{B.4}$$

*Proof —* Let $h > 0$ and consider $g$ defined on $[0,1)$ by

$$g(h) = \int_h^1 x^{\alpha-1}(1-x)^{\beta-1}\left(x - \frac{\alpha}{\alpha+\beta}\right)dx \tag{B.5}$$

Since $\alpha > 1$, the function $g$ is $C^1([0,1))$ and

$$g'(x) = -x^{\alpha-1}(1-x)^{\beta-1}\left(x - \frac{\alpha}{\alpha+\beta}\right)$$

We have
$$g(0) = \frac{\Gamma(\beta)\Gamma(1+\alpha)}{\Gamma(\alpha+\beta+1)} - \frac{\alpha\Gamma(\alpha)\Gamma(\beta)}{(\alpha+\beta)\Gamma(\alpha+\beta)} = 0$$
$$g'(0) = 0$$

Subsequently $g(h) = o(h)$.

Let us now compute the Newton ratio of $A$ in $0$

$$\frac{A(h) - A(0)}{h} = \frac{\int_h^1 x^\alpha (1-x)^{\beta-1} dx}{h \int_h^1 x^{\alpha-1} x^{\alpha-1}(1-x)^{\beta-1} dx} - E$$

$$= \frac{g(h)}{h \int_h^1 x^{\alpha-1}(1-x)^{\beta-1} dx}$$

$$= \frac{o(h)}{h\mu([h,1])}$$

$$= \frac{o(1)}{\mu([h,1])} \to 0 \text{ (as } h \to 0)$$

□

**Lemma 10.** *Let $\mu$ be a Beta distribution associated to a point in regions $U$ or $D$ of the EZ-square (that is $\mu = \text{Beta}(\alpha, \beta)$ with $\alpha < 1$). Then $A$ is not differentiable in $0$. Its curve has a vertical tangent in $0$.*

*Proof* — Let $h > 0$. Consider $g$ defined on $[0,1)$ by (B.5) and $\Psi(x) = x^{1-\alpha} g(x)$. We have $1 - \alpha > 0$ and $\Psi$ is $C^1$ on the *open* set $(0,1)$.

$$\Psi'(h) = (1-\alpha)h^{-\alpha}g(h) + h^{1-\alpha}g'(h) = (1-\alpha)h^{-\alpha}o(h) - (1-h)^{\beta-1}(h-E)$$

We have $\Psi(0) = 0$ and $\Psi'(0) = E$. Therefore $\Psi(h) = Eh + o(h)$, hence:
$$g(h) = h^\alpha E + o(h^\alpha)$$

Now consider
$$\frac{A(h) - A(0)}{h} = \frac{g(h)}{h \int_h^1 x^{\alpha-1}(1-x)^{\beta-1} dx}$$

$$= \frac{h^\alpha E + o(h^\alpha)}{h \int_h^1 x^{\alpha-1}(1-x)^{\beta-1} dx}$$

$$= \frac{E + o(1)}{h^{1-\alpha} \int_h^1 x^{\alpha-1}(1-x)^{\beta-1} dx}$$

It follows the Newton coefficient tends toward $+\infty$ when $h$ tends toward $0$. Thus $A$ is not differentiable in $0$. □

We are left to investigate the case of the segment $(\Omega_3\Omega_6)$ of the *EZ*-square, that is $\alpha = 1$. In this case the differentiability of $A$ at $0$ will be the consequence of a more general result showing that unveils a distinctive property (linearity) of $A$ on this segment.

**Lemma 11.** *Consider a two parameter Beta distribution associated to a point on the segment $(\Omega_3\Omega_6)$ of the EZ-square (that is $\alpha = 1$). Then $A$ is linear:*

$$A(\theta) = Z\theta + E \qquad (B.6)$$

*Proof* — Let $f_{1,\beta}$ be the two parameter Beta density. The Average probability of damage $A$ can indeed be computed explicitly

$$A(\theta) = \frac{\int_\theta^1 x(1-x)^{\beta-1}dx}{\int_\theta^1 (1-x)^{\beta-1}dx} = \frac{\frac{(\beta\theta+1)(1-\theta)^\beta}{\beta(\beta+1)}}{\frac{(1-\theta)^\beta}{\beta}} = \frac{1}{\beta+1} + \frac{\beta}{\beta+1}\theta$$

The result is derives from $Z = \frac{\beta}{\beta+1}$ and $E = \frac{\alpha}{\beta+1}$ with $\alpha = 1$. □

The differentiability of $A$ reported in Lemmas 9, 10, and 11 is represented in Figure B.5 using blue, green and red representations, respectively.

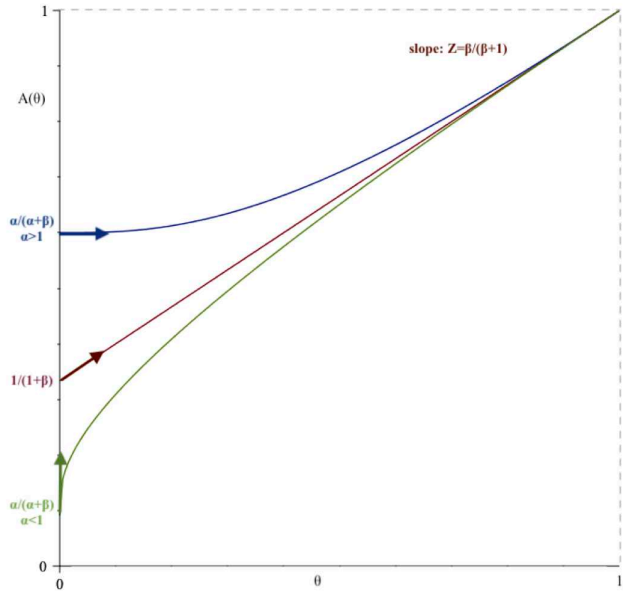

Figure B.5: The function $A$ for a given $\beta$ and various values of $\alpha$.

### Appendix C. A Guide to the Regions of *EZ*-Square

A set $\mathcal{E}$ is *infinite* if there a subset $\mathcal{F}$ distinct from $\mathcal{E}$ that is in bijection with $\mathcal{E}$. This means there is "one-to-one correspondence" that pairs each element in $\mathcal{E}$ with a counterpart element in $\mathcal{F}$, guaranteeing no element is left unpaired. This, though potentially counter intuitive initially, indeed forms the actual the definition of an infinite set.

Taking into consideration the set of real numbers, we can unveil an inherent bijection with the bounded interval $(-\pi/2, \pi/2)$. To envision this, envisage sketching a circle and subsequently introducing a tangent line to this circle. Picking a point on this line, draw a line from this point to the circle's center, and gauge the ensuing angle. This process effectively pairs any point on the line to a corresponding angle within the restricted interval $(-\pi/2, \pi/2)$. In more precise terms, the mapping $x \mapsto \arctan(x)$ associates every real number with an element within the *bounded* interval $(-\pi/2, \pi/2)$. Consequently, the outcome is a compression of the vast continuum of real numbers into the confines of a bounded open interval.

The parameters $\alpha$ and $\beta$ of the Beta distribution both lie in $(0, +\infty)$. The mapping $\varphi$ defined (40) is a bijection between $(0, +\infty) \times (0, +\infty)$ and $(0,1) \times (0,1)$. This bijection is continuous and the inverse map $\varphi^{-1}$ is continuous as well, thus making $\varphi$ an homeomoprhism, preserving the topological structure of the spaces involved. This transformation allows us to parametrize any Beta distribution using a new pair of parameters: $E$ and $Z$. These new parameters have a better economic significance within the context of our insurance problem. Furthermore, they are confined within a bounded interval, which opens the door to the possibility of generating graphs that facilitate a comprehensive and insightful analysis.

The $EZ$-square can be decomposed in four regions delimited by the second diagonal $]\Omega_3, \Omega_6[$ $(Z = 1 - E)$ and the horizontal perpendicular bisector $]\Omega_1, \Omega_4[$ $(Z = \frac{1}{2})$. See Figure C.9.

- The region **D** is the open set delimited by Triangle $(\Omega_0 \Omega_3 \Omega_4)$. In this region, the Beta distributions are Decreasing (see Figure C.6).

- The region **I** is the open set delimited by Triangle $(\Omega_0 \Omega_1 \Omega_6)$. In this region, the Beta distributions are Increasing (see Figure C.6).

- The region **U** is the open set delimited by Trapezoid $(\Omega_0 \Omega_4 \Omega_5 \Omega_6)$. In this region, the Beta distributions are U-shaped (see Figure C.8).

- The region **A** is the open set delimited by Trapezoid $(\Omega_0 \Omega_1 \Omega_2 \Omega_3)$. In this region, the Beta distributions are Arched (see Figure C.7).

- The line segment $]\Omega_0, \Omega_1[$ corresponds to Beta Distributions representing markets where the risk is increasing and modeled by the power distributions of the form $f(x) = \frac{1}{k+1} x^k$ with $k > 0$.

- The line segment $]\Omega_0, \Omega_3[$ corresponds to Beta Distributions representing markets where the risk is decreasing and modeled by the power distributions $f(x) = \frac{1}{k+1} x^k$ with $k > 0$.

- The line segment $]\Omega_0, \Omega_4[$ corresponds to Beta Distributions representing markets where the risk is decreasing with a pole in 0, modeled by a polar power distributions $f(x) = \frac{1}{k+1}(1-x)^k$ with $k \in (0,1)$.

- The line segment $]\Omega_0, \Omega_6[$ corresponds to Beta Distribution representing markets where the risk is increasing with a pole in 1, modeled by a polar power distributions $f(x) = \frac{1}{k+1} x^k$ with $k \in (0,1)$.

- Finally, the center of the square $\Omega_0 = (\frac{1}{2}, \frac{1}{2})$ corresponds to the uniform distribution (Beta with $\alpha = \beta = 1$).

The sets **D**, **I**, **U**, **A**, $]\Omega_0, \Omega_1[$, $]\Omega_0, \Omega_3[$, $]\Omega_0, \Omega_4[$, $]\Omega_0, \Omega_6[$ and $\{\Omega_0\}$ form a partition of the $EZ$-square $(0,1)^2$.

All Beta distributions belong to $\mathcal{M}_2$ regardless of the parameters and the corresponding regions. However, only Beta distributions in **A** and **I** seem to be members of $\mathcal{M}_3$. This is based on the subsequent conjecture, which is substantiated by numerical validation [11].

**Conjecture 1.** *Let $\mu$ be a Beta distribution with $\alpha > 1$ (i.e. $Z \geq 1 - E$) then A is strictly convex.*

We have
$$\frac{\partial \operatorname{Var}(E, Z)}{\partial Z} = -\frac{E(1-E)^2}{[1 - E(1-Z)]^2} < 0$$

The following result thus is true.

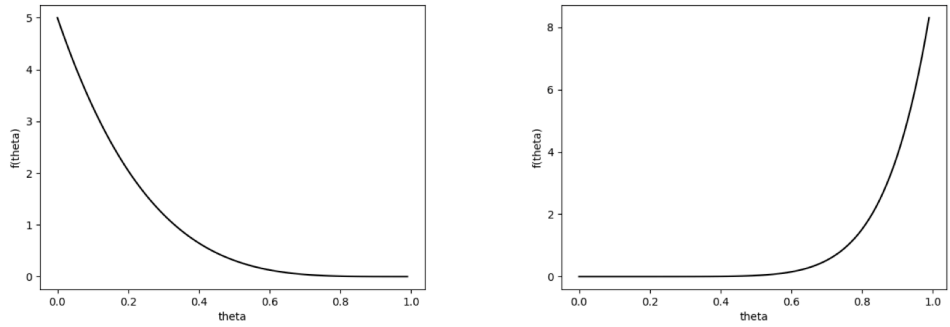

Figure C.6: Examples of a Beta distribution associated to a point in regions **D** (left) and **I** (right) of the $EZ$-square

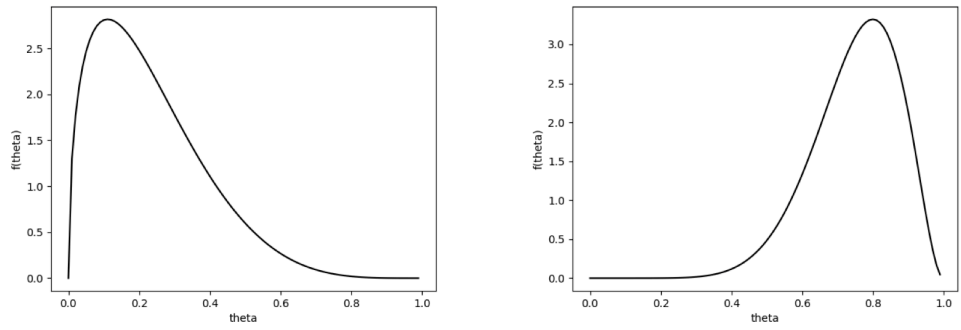

Figure C.7: Examples of a Beta distribution associated to points in region **A** of the $EZ$-square

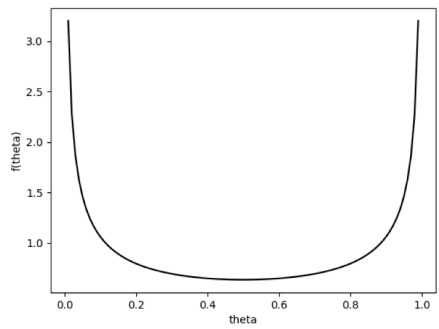

Figure C.8: Example of a Beta distribution associated to a point in the region **U** of the $EZ$-square

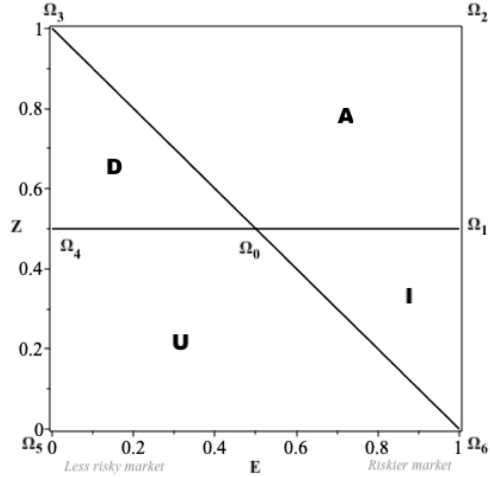

Figure C.9: The $EZ$-square

**Fact 7.** *For any point $(E, Z)$ located in the $EZ$-square $(0,1) \times (0,1)$, the variance (of the Beta distributed random variable $\zeta$) is a decreasing function of $Z$ and tends to zero when $Z$ tends to one.*

Consider a sequence of random variables $(\zeta_n)_{n \in \mathbb{N}}$ following a Beta distribution in the $EZ$ square defined by a given $E \in (0,1)$ and $Z_n \in (0,1)$ where $\lim_{n \to +\infty} Z_n = 1$. Then $(\zeta_n)_{n \in \mathbb{N}}$ converges in law to a random variable following the Dirac distribution $\delta_E$. This scenario represents a situation where *all* agents share the same type $E$. Because the insurer has knowledge of the distribution, it operates under complete information. The upper horizontal boundary of the $EZ$-square, represented by the segment $(\Omega_3 \Omega_2)$ corresponds to cases where Dirac masses are present and they are *not* represented by Beta distributions. Following [12], an extension by continuity could be undertaken to include scenarios with Dirac masses where $Z = 1$, thereby including the upper boundary of the $EZ$-square within the analysis.

We call a *vertical segment* or *section* within the $EZ$-square a segment that is parallel to $(\Omega_5 \Omega_3)$, meaning it is parallel to the $Z$-axis. For a given positive $E \in (0,1)$, it was established that as $Z$ decreases, the variance decreases as well.

**Remark 4.** *From the results contained in [12], that increasing the variance in a two-parameter Beta distribution is equivalent to implementing a mean-preserving-spread. This process results in a change in the shape of the Beta distribution as $Z$ decreases.*

**Appendix D. Implementation and Computational Details**

Numerical simulations were conducted using Python scripts. The scripts were developed in Python version 3.7.6, using the numerical library Numpy version 1.18.1. The computations were executed using the high-performance computing (HPC) resources from the Mesocentre computing facility[11] at CentraleSupélec and École Normale Supérieure Paris-Saclay.

When there is no regulation on indemnity, the maximization of profit with respect to $\theta$ and $r$ was achieved using a dual annealing algorithm. In cases where indemnity regulation was imposed, the maximization of profit with respect to $\theta$ was performed using the Bounded Brent method.

---

[11]This facility is partially funded by the CNRS and the Région Île-de-France.

The primary routines can be found in the files `onecontract.py` for the main analysis, `fig4.py`, `fig5.py`, and `fig6.py` produce 3, and 4, `table1.py` produces Table 1 and `conjecture.py` numerically verifies Conjecture 1 on the convexity of function $A$. These scripts are released under the Creative Commons Attribution-NonCommercial 4.0 International License (CC-BY-NC-4.0). They can be accessed at [11], allowing for result reproducibility, inspection, and potential improvement.

We now complete the non-dimensionalization process. The non-dimensionalized wealth $w$, loss $l$, indemnity $r$ and price $p$ are respectively defined by $w = W/W_0$, $l = L/W_0$, $r = R/W_0$ and $p = P/W_0$. The market segmentation parameter $\theta$ is already non-dimensional. The non-dimensionalized willingness-to-pay $\bar{p}(\theta, r)$, with parameters $l$ and $\rho$, should be equal to $\bar{P}(\theta, rW_0)/W_0$ with parameters $lW_0$ and $\rho W_0$. Equation (8) yields:

$$\bar{p}(\theta, r) = \frac{1}{\rho} \times \ln\left(\frac{\theta e^{\rho l} + (1-\theta)}{\theta e^{\rho(l-r)} + (1-\theta)}\right) \tag{D.1}$$

This function is parameterized by $\rho$. It is easy to see that when $\rho$ tends to infinity, it tends to $r$, that is, when agents are infinitely risk-averse, they are ready to pay $r$ for the insurance contract regardless of their probability of damage $\theta$. The non-dimensional profit $\pi$ is defined by $\pi(\theta, r) = \Pi(\theta, rW_0)/W_0$. Equation (9) yields:

$$\pi(\theta, r) = [\bar{p}(\theta, r) - r \times A(\theta)][1 - F(\theta)] \tag{D.2}$$

**Remark 5.** *It might appear surprising at first glance to divide the dimensional profit $\Pi$ of the insurer by the agent's wealth $W_0$ to obtain the non-dimensional profit $\pi$. Actually, one can divide the dimensional variables by any quantity with the dimension of a currency unit in order to achieve dimensionless variables. This choice doesn't impact the determination $(\theta^*, r^*)$. Furthermore, once the non-dimensional value of $\pi(\theta^*, r^*)$ is found, it is re-scaled using the constant used, here $W_0$. Therefore, an alternative choice to $W_0$ could have been made, which would have required adjustments to the formulas but wouldn't have altered the final results.*

The non-dimensional surplus sp is defined by $\mathrm{sp}(\theta, r) = \mathrm{SP}(\theta, rW_0)/W_0$. Equation (10) yields:

$$\mathrm{sp}(\theta, r) := \int_\theta^1 (\bar{p}(x, r) - \bar{p}(\theta, r))\, d\mu(x) \tag{D.3}$$

The non-dimensional social welfare sw is defined by $\mathrm{sw}(\theta, r) = \mathrm{SW}(\theta, rW_0)/W_0$. Equation (10) yields:

$$\mathrm{sw}(\theta, r) := \pi(\theta, r) + \mathrm{sp}(\theta, r) = \int_\theta^1 (\bar{p}(x, r) - rx) d\mu(x) \tag{D.4}$$

Finally, let us introduce $\xi_r(\rho) = \Xi_{rW_0}(\rho/W_0)$. Equation (25) yields

$$\xi_r(\rho) = \frac{e^{-\rho l}(1 - e^{-\rho r})}{\rho r} \tag{D.5}$$

The optimal value of $\theta$ that maximizes the profit function $\theta \mapsto \Pi(\theta, R)$ for a given $R \in (0, L)$ has been denoted as $\theta_R^*$. We will denote $\vartheta_r^*$ the point where the function $\theta \mapsto \pi(\theta, r)$ is maximized. We have $\vartheta_r^* = \theta_{rW_0}^*$. For the sake of simplicity in our notations, we will use $\theta_r^*$ instead of $\vartheta_r^*$, which is an abuse of notation.

Since the dimensionless model is formally equivalent to the dimensional one, all equations are identical; the dimensional variables of the model presented in the previous section are simply replaced by their non-dimensional counterparts.